\documentclass[a4paper,12pt]{article}
\pdfoutput=1 
\RequirePackage{amsthm,amsmath,amsfonts,amssymb}
\RequirePackage[authoryear]{natbib}
\RequirePackage[colorlinks,citecolor=blue,urlcolor=blue]{hyperref}
\RequirePackage{graphicx}

\usepackage[caption=false]{subfig}
\usepackage{float}
\usepackage[dvipsnames,table,xcdraw]{xcolor}
\usepackage{blindtext}
\usepackage{hyperref}
\usepackage{ulem}
\usepackage{setspace}
\usepackage{verbatim}
\immediate\write18{texcount -tex -sum  \jobname.tex > \jobname.wordcount.tex}
\usepackage{sectsty}

\usepackage[english]{babel}

\usepackage[legalpaper, margin=1in]{geometry}
\usepackage[utf8]{inputenc}

\usepackage{setspace}   
\usepackage[affil-it]{authblk} 
\usepackage{etoolbox}
\usepackage{lmodern}

\providecommand{\keywords}[1]
{
  \small	
  \textbf{\textit{Keywords---}} #1
}

\makeatletter
\patchcmd{\@maketitle}{\LARGE \@title}{\fontsize{16}{19.2}\selectfont\@title}{}{}

\sectionfont{\fontsize{14}{15}\selectfont}
\subsectionfont{\fontsize{13}{15}\selectfont}

\makeatother

\title{\textbf{Pairwise Nonlinear Dependence Analysis of Genomic Data}}

\author[1]{Siqi Xiang}
\author[1]{Wan Zhang}
\author[2,3,4]{Siyao Liu}
\author[2,3,4]{Katherine A. Hoadley}
\author[2,3,4]{Charles M. Perou}
\author[1]{Kai Zhang}
\author[1,2,4]{J. S. Marron}

\affil[1]{Department of Statistics and Operations Research, University of North Carolina at Chapel Hill}
\affil[2]{Lineberger Comprehensive Cancer Center, University of North Carolina at Chapel Hill}
\affil[3]{Department of Genetics, School of Medicine, University of North Carolina at Chapel Hill}
\affil[4]{Curriculum in Bioinformatics and Computational Biology, University of North Carolina at Chapel Hill}
\affil[ ]{\textit{\href{mailto:xsiqi@email.unc.edu}{xsiqi@email.unc.edu}, \href{mailto:wan.zhang@unc.edu}{wan.zhang@unc.edu},
\href{mailto:siyao@email.unc.edu}{siyao@email.unc.edu},
\href{mailto:hoadley@med.unc.edu}{hoadley@med.unc.edu},
\href{mailto:cperou@med.unc.edu}{cperou@med.unc.edu}, \href{mailto:zhangk@email.unc.edu}{zhangk@email.unc.edu}, \href{mailto:marron@unc.edu}{marron@unc.edu}}}

\date{}

\begin{document}
\maketitle
\pagestyle{plain}

\begin{abstract}
In The Cancer Genome Atlas (TCGA) data set, there are many interesting nonlinear dependencies between pairs of genes that reveal important relationships and subtypes of cancer. Such genomic data analysis requires a rapid, powerful and interpretable detection process, especially in a high-dimensional environment. We study the nonlinear patterns among the expression of pairs of genes from TCGA using a powerful tool called Binary Expansion Testing. We find many nonlinear patterns, some of which are driven by known cancer subtypes, some of which are novel.
\end{abstract}


\section{Introduction}

A leading cause of death in the world is cancer. A lot of cancer research is currently analyzing larger and larger data sets. In this paper, we focus on The Cancer Genome Atlas (TCGA) (\citeyear{cancer2012comprehensive}), which is a particularly important and comprehensive data set to study cancer biology and genomics. It is a publicly available genomics data resource that seeks to understand several types of cancer by collecting multiple diverse data over many people. In that set, an important data type is gene expression, and more specifically, RNA-seq data. A biologically useful task of modern genomics data analysis is detecting the dependency patterns among gene expression. Conventional approaches to dependency, such as Pearson's, Spearman's rank, or Kendall's rank correlation coefficients, target linear dependence. The focus of this paper is a much deeper investigation of nonlinear dependence in TCGA breast cancer data. The dependence of two such genes was an example shown in \cite{zhang2019BET}. Here we carry the applied analysis much further by a detailed study of all pairs of genes. Furthermore, we take the analysis even deeper by studying gene dependence within subtypes as well. We show nonlinear dependence plays a much stronger role than previously imagined by finding $167,173$ interesting nonlinear dependent pairs in the full data set. As seen in Table \ref{compare} below, only $36.3\%$ of these significant pairs were discovered by using the classical Hoeffding’s D statistic.

\begin{figure}[]
\centering
{%
\resizebox*{4cm}{!}{\includegraphics[width=1\textwidth,height=1\textheight,keepaspectratio]{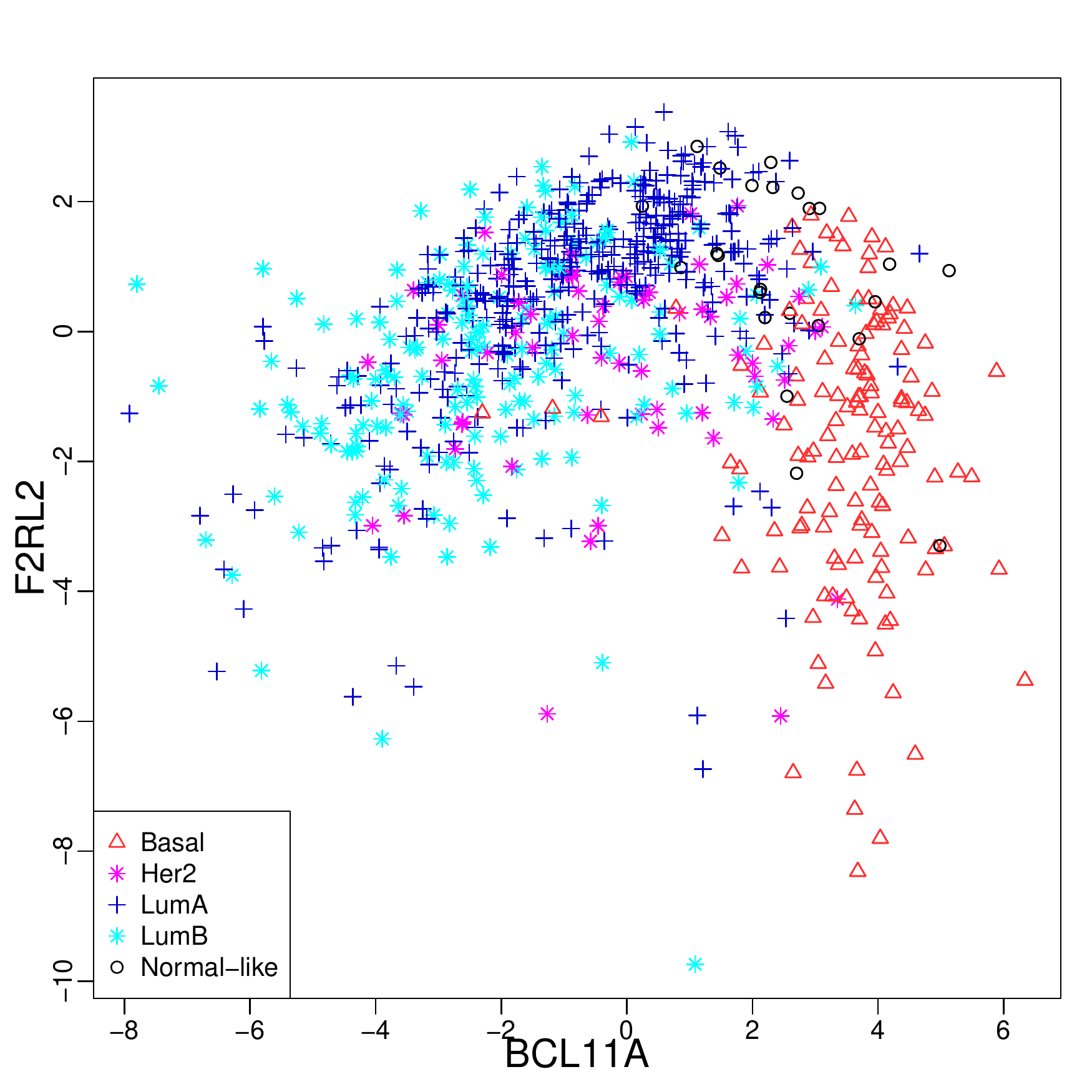}}}\hspace{5pt}
{%
\resizebox*{4cm}{!}{\includegraphics[width=1\textwidth,height=1\textheight,keepaspectratio]{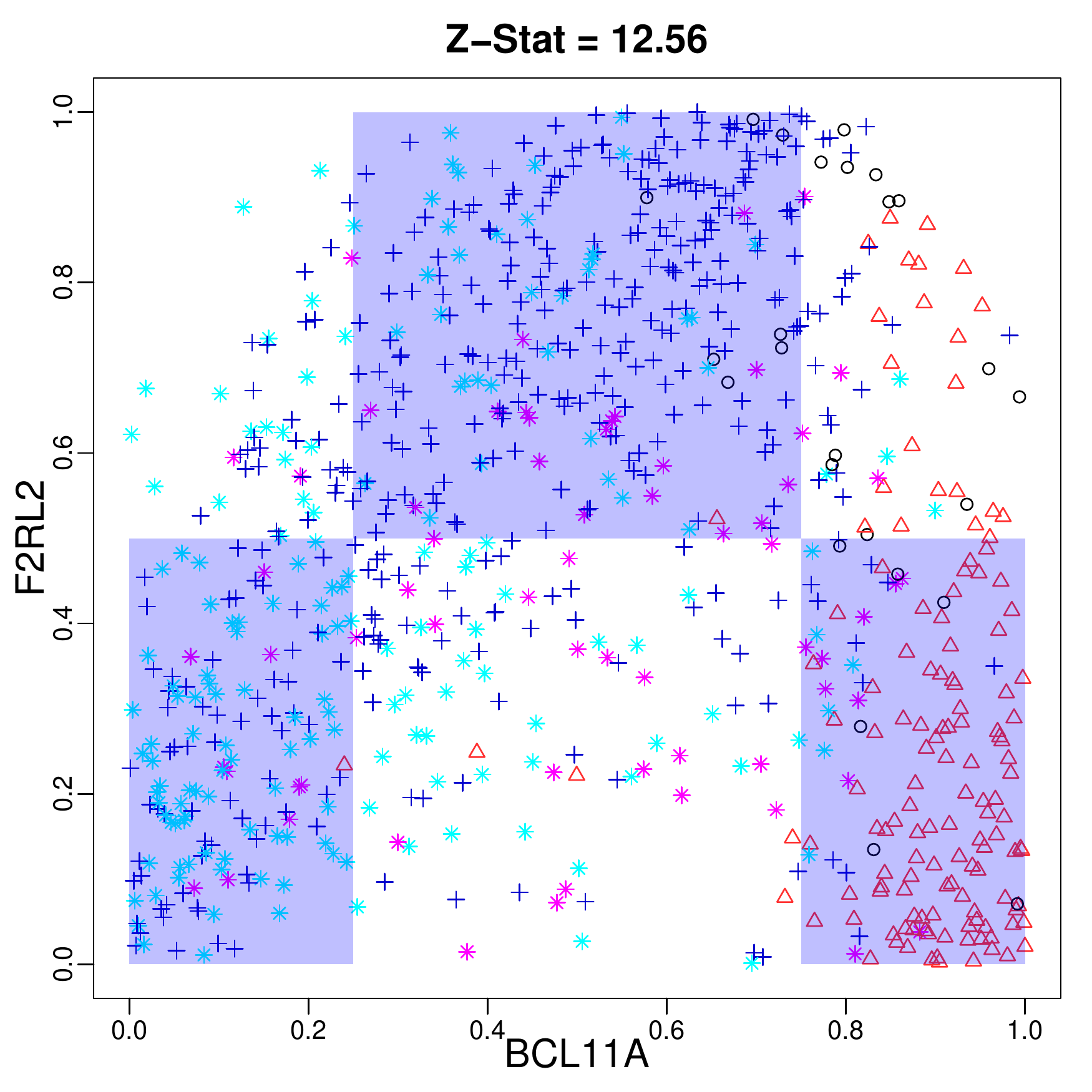}}}
\caption{Left: The scatter plot comparing expression of two genes in TCGA breast cancer data in the normalized log count scale; Right: The scatter plot of the same two genes using the copula transformation with the nonlinear dependence pattern from BET. Strong statistical significance is indicated by the BET Z-statistic of 12.84. These pairwise genes exhibit an interesting nonlinear dependence pattern which is explained by showing the breast cancer subtypes.} \label{intro-figure1}
\end{figure}

An example of the expression of two genes with strong and important nonlinear dependence that is not discoverable by linear methods is shown in the left panel of Figure \ref{intro-figure1}. That is a scatter plot of expression for the genes BCL11A and F2RL2. For this pair of genes, the Pearson correlation coefficient is -0.0069, Spearman's $\rho$ is 0.088, and Kendall's $\tau $ is 0.085. These correlation coefficients are all very close to zero, suggesting no linear correlation between these two genes, as is also visually apparent. However, there is clear nonlinear dependence. This dependence is explained from a biological viewpoint by labeling with commonly used breast cancer subtypes, which were originally discovered by clustering some carefully selected genes in \cite{perou2000molecular}. As shown by the colors and symbols in Figure \ref{intro-figure1}, the decreasing part on the right (suggesting negative correlation) is mainly caused by the Basal (\textcolor{red}{$\bigtriangleup$}) subtype observations. For gene network considerations, the increasing part on the left, driven by the Luminal A (\textcolor{blue}{$+$}) and Luminal B (\textcolor{cyan}{$\ast$}) subtypes, suggest a positive correlation. This important biological refinement of network analysis is unavailable from classical gene network approaches based on conventional correlation measures. This nonlinear dependence pattern has been discovered by Binary Expansion Testing (BET) proposed by \citet{zhang2019BET}. The right panel of Figure \ref{intro-figure1} is the corresponding BET diagnostic plot explained in Section \ref{2.1}.

The data set studied in this paper consists of gene expression features of TCGA Lobular Freeze breast cancer data from \cite{ciriello2015comprehensive}, containing 16615 genes. Note that the total number of pairwise comparisons of genes is $ {16615\choose 2} = 138,020,805$. Human visualization of all of these scatter plots is intractable \citep{Sun2014ThingsInOrder}. An interesting early approach to this is Tukey's scagnostics \citep{wilkinson2005graph}. The large number of pairs motivates a computationally efficient method for investigating pairwise dependence. Gene expression studies have revealed a large number of linear dependencies between genes. In particular, in our BET analysis, we discover $10,110,787$ pairs of (statistically significant using Bonferroni multiple comparison adjustment) dependencies between genes. A large number of these are well understood. This paper takes genomics in a new direction by investigating nonlinear dependencies of the type shown in Figure \ref{intro-figure1}.

There are several current approaches to studying nonlinear dependence. An early measure was Hoeffding's D \citep{Hoeffding1948NonparaIndTest} which is not particularly powerful in the direction of nonlinear dependence \citep{zhang2019BET}. More recently, Sz\'{e}kely et al. have proposed the more powerful (in the direction of nonlinear dependence) method of distance correlation \citep{szekely2007DistCorr,MR3055745}. Another more powerful approach is the $k$-nearest neighbor mutual information (KNN-MI) algorithm \citep{Kraskov2004MutualInfo,Kinney2014MutualInformation}, which focuses on mixtures of gaussian distributions. While these methods are beneficial for discovering nonlinear dependence, they are less suitable for extensive genomic studies for three reasons. First, they are not efficient for large-scale computation problems such as TCGA data (tens of thousands of genes and hundreds of samples). Second, they still face some power loss in the direction of nonlinear dependence, as noted using simulation studies in Section 6 of \citet{zhang2019BET}. Third, there is less immediate interpretation of the type that is available from BET. 

To demonstrate the relatively slow execution time needed for the three methods above, we compare their calculation speed with BET. The running time for testing all pairs of a randomly selected set of 100 genes is shown in Table \ref{table1}. We use the default setting for each algorithm.

\begin{table}[!h]
\centering
\caption{The running time of the pairwise comparison of 100 genes. The more powerful nonlinear detection methods are orders of magnitude slower than BET.}
\label{table1}
\begin{tabular}{@{}lrrrrc@{}}
\hline
Algorithms& \multicolumn{1}{c}{BET}
& \multicolumn{1}{c}{Hoeffding's D}
& \multicolumn{1}{c}{Distance Correlation} 
& \multicolumn{1}{c}{KNN Mutual Information}  \\
\hline
Times   & 8.96 secs & 24.05 secs & 17.51 mins  & 4.91 hours  \\
\hline
\end{tabular}\label{table1}
\end{table}

Table \ref{table1} shows that BET is around three times faster than Hoeffding's D which has less power in the direction of nonlinear dependence, as seen in Section \ref{3.2}. It also shows that BET provides computational speed that is several orders of magnitude faster than either distance correlation or KNN mutual information in the context of dependence testing of high-dimensional data such as TCGA. Furthermore, nonlinear dependence can arise in many forms. As mentioned above, another advantage of BET is that it gives additional information on the form of nonlinear dependence, as illustrated in Figure \ref{Figure-2}.

This paper is organized as follows. Section \ref{2} describes the main idea of the Binary Expansion Testing (BET) algorithm. Section \ref{3} details the BET analysis of TCGA data set, revealing several interesting nonlinear patterns. Section \ref{4} studies the validation of some surprising TCGA results using an independent genomics data set. Section \ref{5} concludes the article.

\section{Binary Expansion Testing for Nonlinear Dependency Detection}\label{2}

BET is a recent and innovative approach for dependence testing that is powerful for detecting pairwise nonlinear dependence. Furthermore, it provides a computationally fast investigation process for large-scale data sets, as shown in Table \ref{table1}. Finally, BET gives a clear interpretation for some specific nonlinear dependence patterns. These patterns are formally introduced in Section \ref{2.1}. The BET algorithm and inference are given in Section \ref{2.2}.

\subsection{BET Dependence Patterns}\label{2.1}

\begin{figure}[]
\centering
\includegraphics[width=0.55\textwidth,height=0.55\textheight,keepaspectratio]{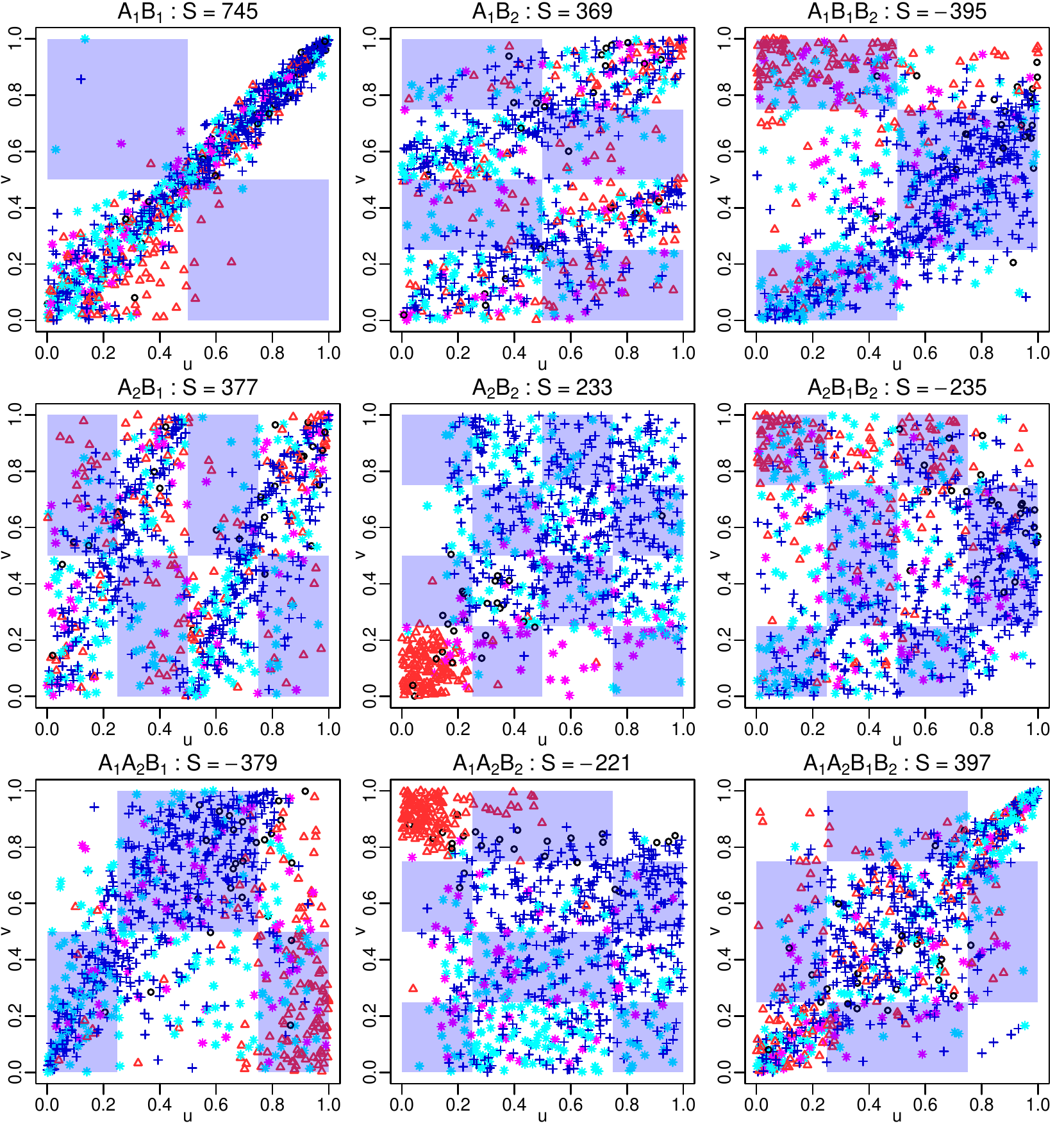}
\caption{The first nine low-resolution dyadic binary interaction designs (BIDs) used by BET and one example pair of expression data shown for every BID, with $S$ indicating the difference of counts in white and blue regions. Each BID is aimed at detecting a particular dependence relationship between two $[0,1]$ uniform random variables. Size of $S$ indicates the strength of nonlinear dependence in each BID. A number of biologically relevant patterns are shown.}
\label{Figure-2}
\end{figure}

The first step of BET is the copula transform of the bivariate distribution to the unit square $[0,1]^2$ using a marginal probability integral transformation of each variable in the pair, as detailed in Section \ref{2.2}. The key idea of BET is to partition the unit square into different patterns that indicate interesting types of dependence in terms of counts, i.e., densities of observations, in different regions. For fast computation, these patterns are dyadic in nature. The first few of these are shown in Figure \ref{Figure-2} where unions of blue blocks represent one region and white blocks are the alternative. Each partition pattern is called a \textit{Binary Interaction Design (BID)} by \citet{zhang2019BET}. In our genomics data analysis,we consider the nine dependence BIDs shown in Figure \ref{Figure-2}. Each BID corresponds to one BET dependence pattern. If there is no dependence between $U$ and $V$, the observations are randomly distributed in $[0,1]^2$ . Dependence patterns are reflected by significant differences between these blue and white region counts of points (density of observations). For each form of BID, the difference is called $S$ (the \textit{symmetry statistic}). These counts are tested against the null hypothesis of no difference between them. The value of $S$ is given for each BID in Figure \ref{Figure-2}. Notice that $S$ is positive if the white part is denser than the blue and negative otherwise. For example, the white region in the upper-left panel of Figure \ref{Figure-2} (the label $A_1B_1$ of that BID will be explained later) contains more points and captures a monotone upward dependence, which corresponds to the large positive $S = 745$. Had the linear dependence in the data been downward, there would have been greater density in the blue region, and $S$ would be negative. An example where the blue region is more dense appears in the BID in the upper-right panel, where $S = -395$. Each dependence pattern in Figure \ref{Figure-2} is illustrated using a pair of genes that strongly exhibits the corresponding BID, particularly the pair with the maximal absolute value of $S$.

Figure \ref{Figure-2} shows that BET captures several different nonlinear relationships for strongly dependent pairs of genes. Specific genes in these pairs are listed in Table \ref{table2}, where Gene 1 is on the horizontal axis and Gene 2 is on the vertical axis. As noted above, the symmetry statistic $S$ for each BID is far from 0, indicating that strong dependence exists. For example, the middle-left panel highlights a surprising bimodal pattern that we will discuss in detail in Section \ref{bio_pat} and \ref{4.1}. The bottom-left panel indicates dependence roughly following a downward opening parabola, which is captured by a high density in the blue region with corresponding negative $S$ value. Left and right opening parabolas are captured by the BID in the upper-right, where this pair of genes has a leftward opening parabola. In contrast, the bottom-right panel looks quite different. It is fairly close to linear dependence, but clearly not bivariate normal. There are unusual concentrations in the upper right and the lower left.

\begin{table*}[!h]
\centering
\caption{Gene names for each pair in Figure \ref{Figure-2}}
\label{sphericcase}
\begin{tabular}{@{}lrrrrc@{}}
\hline
Location & Gene 1 ($U$) & Gene 2 ($V$) \\
\hline
top-left   & C17orf81 & C17orf61  \\
top-middle   & RPL24 & RPL9  \\
top-right   & JAM3 & ANKS6  \\
middle-left   & RPL9 & RPL32  \\
center   & PRR15 & CA12  \\
middle-right   & CDH5 & ZNF883  \\
bottom-left   & ANKS6 & DCN  \\
bottom-middle   & CT62 & FAM174A  \\
bottom-right   & PROSC & ASH2L  \\
\hline\label{table2}
\end{tabular}
\end{table*}

The prevalence of these mixture patterns and some of the others frequently turns out to be a consequence of known breast cancer subtypes. In particular, Basal is known to be very distinct. The dependencies highlighted by a number of different BIDs in Figure \ref{Figure-2} are clearly explained by the separation between Basal (\textcolor{red}{$\bigtriangleup$}) and the other subtypes. The two mixture patterns, in the bottom-left and the top-right BIDs, demonstrate how the mixture of Basal and other subtypes can drive such patterns. The great difference between the Basal subtype and the rest also drives the dependency patterns in the center ($A_2B_2$) and the bottom-middle ($A_1A_2B_2$) BIDs. It is important to remember that several BIDs can respond (i.e., have a significant absolute value of $S$) to a given dependence pattern in the data. However, we made the computational choice of only keeping the largest absolute $S$ value for each pair of genes. Hence, the top 200 most significant genes shown in the graphs in Figures \ref{3_1_network} and \ref{3_4_network} may miss some overlapping pairs of genes. Potential future work of interest would be to study the impact of this choice.

From classical statistical viewpoints, such as those based on sparsity and correlation analysis of dependence, this data set has a perhaps surprising amount of nonlinear dependence, as seen from the numbers of significant gene pairs shown in Figure \ref{SummaryOfResults}.

\subsection{BET Algorithm and Inference}\label{2.2}

Next we formally introduce the testing procedure and notations. For each pair of genes, consider the sample data as pairs of variables $(X_1, Y_1)$, ..., $(X_n, Y_n)$. We view these as realizations of two random variables $X$ and $Y$. BET is a fully nonparametric method based on the copula transformation, which is computed from the marginal CDFs. In particular, let $U = F_X(X)$ and $V = F_Y(Y)$ which are uniform on $[0,1]$ and preserve the relative relationship between $X$ and $Y$. Because the CDFs $F_X(X)$ and $F_Y(Y)$ are often unknown in practice, BET approximates them using the empirical CDF. Thus the $i$th observation in the empirical copula is $(\hat{U}_i, \hat{V}_i)$ whose marginal distribution is uniformly distributed on the equally spaced support points $\frac{1}{n},...,\frac{n-1}{n},1$ on $[0,1]$.

A key motivation for BET is that each decimal fraction number in the interval $[0,1]$ has a binary representation. The quick to compute binary white and blue  dyadic subinterval patterns that underlie Figure \ref{Figure-2} are motivated by the useful probabilistic binary expansions of the continuous uniform random variables $U$ and $V$. These binary expansions \citep{independenceKac1959} are $U = \sum_{k=1}^\infty A_k/2^k$ and $V = \sum_{k'=1}^\infty B_{k'}/2^{k'}$, where $A_k \stackrel{iid}{\sim} Bernoulli(1/2)$ and $B_{k'} \stackrel{iid}{\sim} Bernoulli(1/2)$. Similarly, each observation in the empirical copula $\hat{U}_i$ and $\hat{V}_i$ also has a binary expansion: 
$\hat{U}_i = \sum_{k=1}^\infty \hat{A}_{k,i}/2^k$ and $\hat{V}_i = \sum_{k'=1}^\infty \hat{B}_{k',i}/2^{k'}$.
Note that the binary expansion of an observation $\hat{U}_i$ is the binary representation of this number. Thus, $\hat{A}_k$ and $\hat{B}_{k'}$ can be regarded as the 0-1 indicator functions containing the randomness of $\hat{U}_i$ and $\hat{V}_i$, for example, $\hat{A}_1= I(\hat{U}_i \in (1/2,1])$ and $\hat{A}_{k} = I (\hat{U}_i \in  \bigcup_{j=1}^{2^{k-1}} ((2j-1)/2^{k}, 2j/2^{k}])$. 

The blue and white regions underlying the BET statistic are based on the truncation of these binary expansions at some finite depths $d_1$ and $d_2$, respectively, $U_{d_1}=\sum_{k=1}^{d_1}A_k/2^k$ and $V_{d_2}=\sum_{k'=1}^{d_2}B_{k'}/2^{k'}$. The discrete variables $U_{d_1} $and $V_{d_2}$ take on at most $2^{d_1}$ and $2^{d_2}$ values. Hence, there are $2^{d_1+d_2}-1$ binary variables resulting from interactions between $A_k$ and $B_{k'}$. These variables are sufficient statistics to study interesting dependence \citep{zhang2019BET}. In order to present these interaction variables in the form of products and reflect dependence between these products, we use the binary variables $\dot{A}_k = 2A_k-1$ and $\dot{B}_{k'} = 2B_{k'}-1$ to replace  $A_k$ and $B_k$. Thus, the interaction events between $A_k$ and $B_{k'}$ can be written as the products $\dot{A}_k\dot{B}_{k'}$. For example, the events $\{A_1 = 1, B_1 = 1\}$ and $\{A_1 = 0, B_1 = 0\}$ lead to the same interaction event $\{\dot{A}_1\dot{B}_1 = 1\}$. Out of these $2^{d_1+d_2}-1$ interactions, there are $(2^{d_1}-1)(2^{d_2}-1)$ variables of product form $\dot{A}_{k_1}...\dot{A}_{k_r}\dot{B}_{{k'}_1}...\dot{B}_{{k'}_t}$ for some $r, t > 0$. We call these variables \textit{cross interactions}, each of which results from the product of at least one $\dot{A}_k$ and one $\dot{B}_{k'}$ and reflects a BET partition (the BID mentioned earlier) in $[0,1]^2$, i.e., each product results in one type of white and blue regions partition. For example, in Figure \ref{Figure-2}, the BID ${A}_1{B}_1$ represents the cross interaction variable $\dot{A}_1\dot{B}_1$, where the observation $i$ in the white region reflects the event $\{\dot{A}_{1,i}\dot{B}_{1,i}= 1\}$ and in the blue region reflects the event $\{\dot{A}_{1,i}\dot{B}_{1,i}= -1\}$. Thus, the depth parameters $d_1$ and $d_2$ decide the amount and type of BIDs considered in BET. The choice $d_1 = d_2 = 2$ is the right resolution to find the most dependence patterns of biological interest (as the nine BIDs shown in Figure \ref{Figure-2}). We do not consider larger choices of $d$ for two reasons. First, since $d=2$ cuts each interval into quarters and $d=3$ cuts each interval into eighths, most of these patterns from larger depths which can not be seen at depth $d=2$ capture relationships that are not expected to give useful insights into the dependence inherent to gene expression. Second, it would entail a considerable computational cost: when given $d_1 = d_2 = 3$, we have a total of $49$ BIDs including these nine patterns, meaning the computational cost is raised by a factor of more than five. \cite{zhang2019BET} gives more discussion about depth selection. Let $\boldsymbol{a}$ and $\boldsymbol{b}$ denote vectors of length $d_1$ and $d_2$ with 1's at $k_1...k_r$ and $k'_1...k'_t$ respectively and 0's otherwise, thus we can denote the cross interaction $\dot{A}_{k_1}...\dot{A}_{k_r}\dot{B}_{k'_1}...\dot{B}_{k'_t}$ as $\dot{A}_{\boldsymbol{a}}\dot{B}_{\boldsymbol{b}}$. For simplicity of notation, the labels ${A}_{\boldsymbol{a}}{B}_{\boldsymbol{b}}$ in most figures represent the cross interactions $\dot{A}_{\boldsymbol{a}}\dot{B}_{\boldsymbol{b}}$.

Now we define the symmetry statistic $S$ for a given cross interaction $\dot{A}_{\boldsymbol{a}}\dot{B}_{\boldsymbol{b}}$ as the difference of counts in white and blue regions in the corresponding BID. Since the value of the cross interaction of the observation is $1$ in white regions and $-1$ in blue regions, we can calculate $S$  as the sum of the observed binary interaction variables $S = \sum_{i=1}^n \dot{A}_{{\boldsymbol{a}},i}\dot{B}_{{\boldsymbol{b}},i}$. The $U_{d_1} $and $V_{d_2}$ are strongly dependent when the absolute value $|S|$ is far from zero for at least one BID, according to the following fundamental observation of \cite{zhang2019BET}: If $U_{d_1} $and $V_{d_2}$ are independent, the symmetry statistic $S$ satisfies $(S+n)/2 \sim Binomial(n, 1/2)$, for $a \neq 0$ and $b \neq 0$. If the empirical copula transformation is used, we can use $\hat{S} = \sum_{i=1}^n \widehat{\dot{A}}_{\boldsymbol{a},i}\widehat{\dot{B}}_{\boldsymbol{b},i}$ as the symmetry statistic and $(\hat{S}+n)/4 \sim Hypergeometric(n, n/2, n/2)$, for $\boldsymbol{a} \neq 0$ and $\boldsymbol{b} \neq 0$.

The issue of multiple comparisons across BIDs is handled by the Max BET procedure of \cite{zhang2019BET}, at depths $d_1$ and $d_2$. This is described as follows for a given pair of variables. First, we compute all symmetry statistics $S$ with cross interactions for the given $d_1$ and $d_2$. Then we look for the symmetry statistic with the strongest asymmetry and record its p-value and z-statistic $|S|/\sqrt{n}$. Finally, for this pair of variables, we use a Bonferroni adjustment across the cross interactions (BIDs) to obtain the corresponding family-wise error rate p-value for this maximum $|S|$. The dependence relationship is represented by the most significant BID.

Notice that Figure \ref{Figure-2} reveals a reflection property among the BIDs. For example, the BIDs $A_1B_1B_2$ (top-right) and $A_1A_2B_1$ (bottom-left) represent the same relationship up to a switch of the axis positions (i.e., reversal of the roles of the two genes) for the two random variables $U$ and $V$. There are three pairs of such off diagonal reflected patterns. Identifying such pairs results in six BID patterns: five nonlinear and one linearity. This reflection property will be further discussed in Section \ref{3.3}.

\section{Results From TCGA}\label{3}
In this section, we first expand on the data preprocessing of TCGA data set in Section \ref{3.1}. Then we summarize the analysis results in Section \ref{3.2}. Finally, we discuss some specific nonlinear dependency patterns in the last few subsections.

\subsection{Data Preprocessing}\label{3.1}
The RNA-seq gene expression features of TCGA Lobular Freeze breast cancer data set from \cite{ciriello2015comprehensive} contain 16615 genes of 817 primary tumor samples, including five subtypes (proportion in the sample): Basal-like ($16.6\%$), HER2 ($8.0\%$), Luminal A ($50.8\%$), Luminal B ($21.5\%$), and Normal-like ($3.1\%$). Intrinsic breast cancer subtyping was done using the PAM50 classifier \citep{parker2009supervised}.

\begin{figure}[]
\centering
{%
\resizebox*{4cm}{!}{\includegraphics[width=0.5\textwidth,height=0.5\textheight,keepaspectratio]{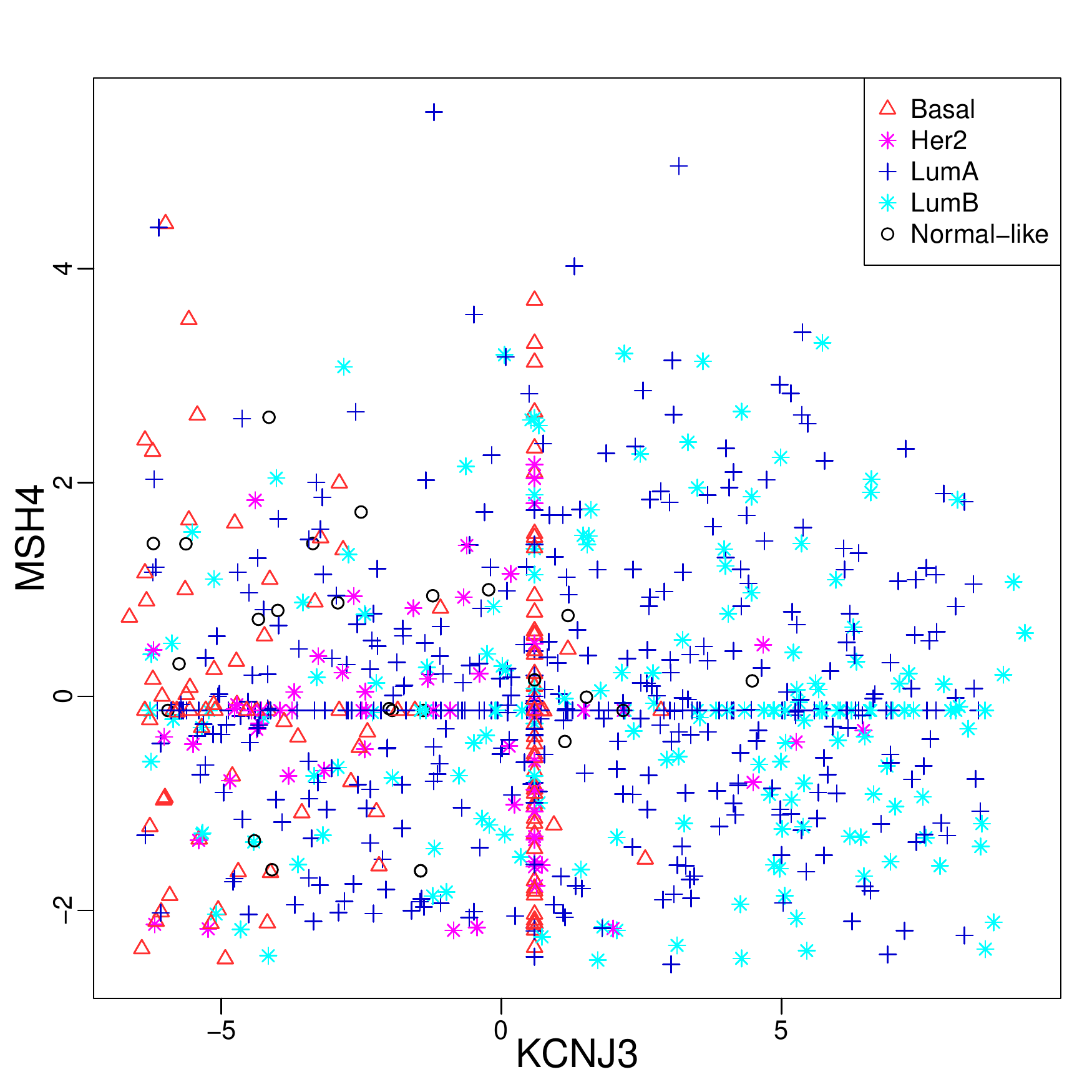}}}\hspace{5pt}
{%
\resizebox*{4cm}{!}{\includegraphics[width=0.5\textwidth,height=0.5\textheight,keepaspectratio]{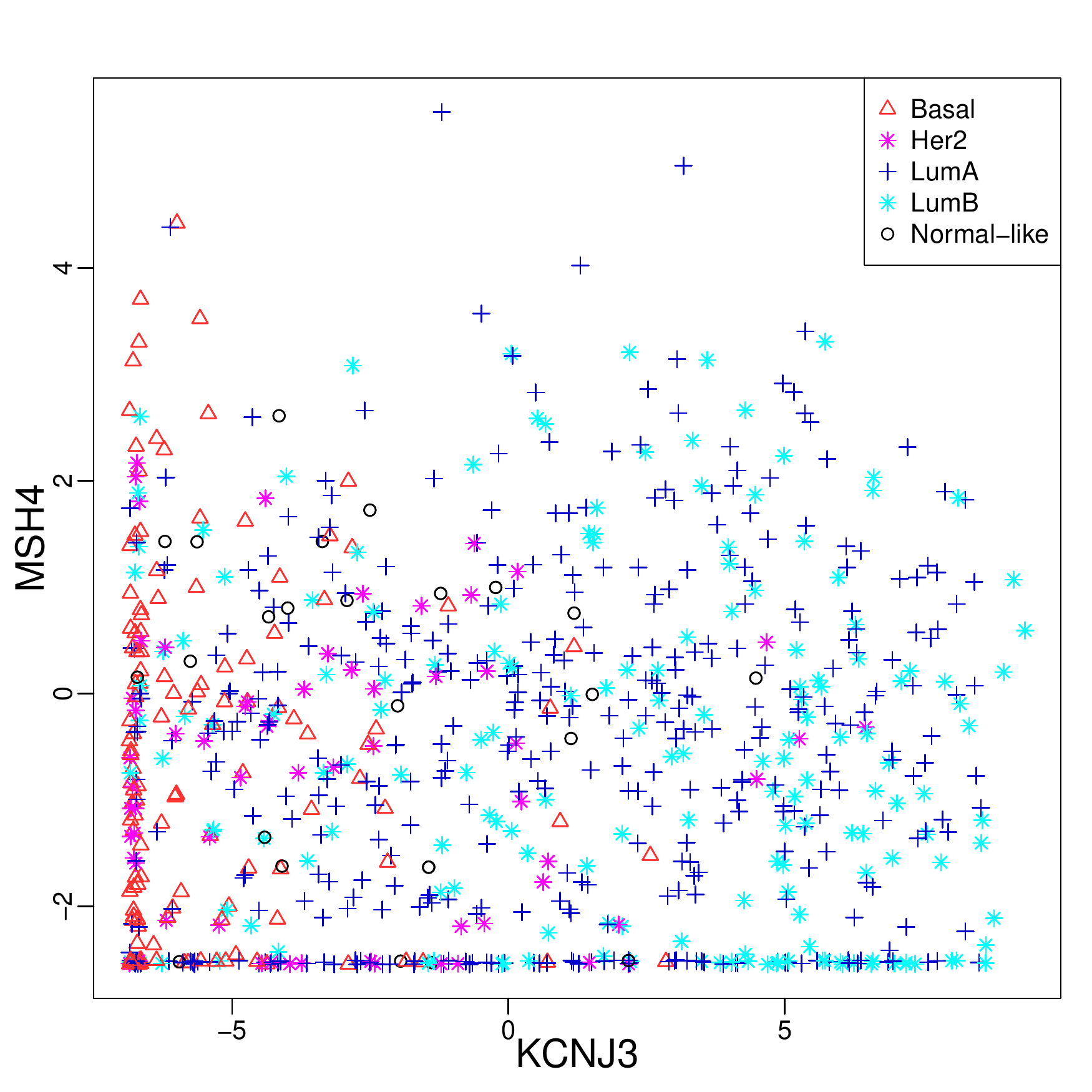}}}\hspace{5pt}
{%
\resizebox*{4cm}{!}{\includegraphics[width=0.5\textwidth,height=0.5\textheight,keepaspectratio]{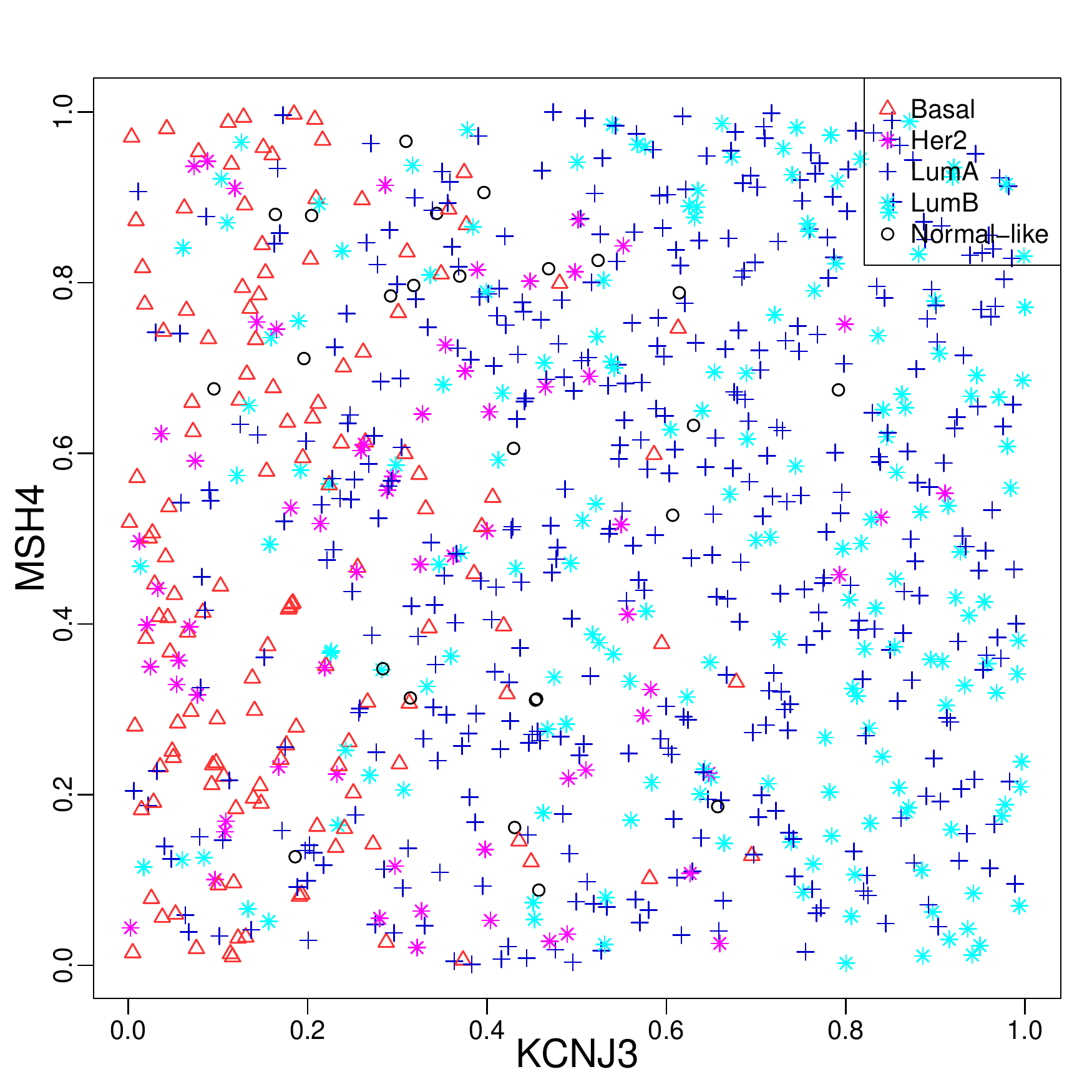}}}
\caption{Left: The scatter plot comparing expression of two genes in the TCGA breast cancer data in the normalized log count scale with zero value imputation by the median; Middle: The scatter plot of the same two genes in the normalized log count scale with the median values reset to the minimum and jittering applied to nonunique values; Right: The scatter plot of the same two genes in the copula distribution scale.} \label{Res_Processing}
\end{figure}

This gene expression data was preprocessed as described in \cite{ciriello2015comprehensive}. This included normalization and logarithms. During this genomics data preprocessing, each sample was normalized to a fixed upper quartile and then $log2$ transformed. Genes with more than $20\%$ zero counts were excluded. Other genes with zero counts had their zeros recorded as missing. A questionable choice made in that preprocessing was to replace these missing values by the median for that gene. However, such data was the beginning of our analysis. The poor consequences of this approach are illustrated in the left panel of Figure \ref{Res_Processing}. In particular, it shows the raw data of a pair of example genes in our TCGA data set, which has many points piled up at the median representing missing values (zero counts). This is an inappropriate way of handling the zeros, because all these data are based on counts, so a zero count represents a small level of gene expression. As an alternative, we first considered moving the median values to be the same as the smallest value. In the case of no zero counts, this is inappropriate because the median corresponds to a non-zero count. A simple fix to this is to take the first non-zero count and leave it at the median. Thus, we address this by setting all but the first of the median values to the minimum value. Some experimentation revealed that this has a minimal impact. The result of this process is shown in the middle panel of Figure \ref{Res_Processing}. This causes many ties for the smallest value. Since BET is based on a copula transform that essentially assumes continuous variables, a large number of ties will cause severe noncontinuity and strongly impact the BET inference. Therefore, these points are spread out in the interval of the minimum and the second unique minimum, using a jitter approach. Specifically, to preserve the ranks, a small random value (uniformly distributed between 0 and the difference between the second unique minimum and minimum) is added to the non-unique minimum observations. Jittering has no impact on the BET significance because the jitter points are within the first $20\%$ of the data (recall that genes with more than $20\%$ missing were excluded). Finally, we apply the empirical copula transformation. See the right panel in Figure \ref{Res_Processing}.

As noted in \citet{parker2009supervised}, these subtypes play a critical role in many aspects of breast cancer. Our analysis in Section \ref{3.3} confirms that the Basal subtype tends to be quite different from the rest. Hence, we also investigate the context of LumA/LumB/Her2, which tends to be dominated by the most numerous subtype LumA. We further investigate the non-LumA group, meaning the union of Her2/LumB, and LumA separately. Hence our analysis focuses on the four different contexts shown in Table \ref{data-contexts} which are chosen to highlight important aspects. Note that the BET analysis focuses on detecting nonlinearity but not subtypes. However, there is a relationship between nonlinearity and subtypes. We set the depth parameters in the BET algorithm $d_1$, $d_2$ to be 2, and only focus on five nonlinear BIDs, as discussed in Section \ref{2.2}. To control for multiple comparisons, in each context, we use the Bonferroni adjustment across genes to modify the BET output p-value of each pair, which has already been adjusted across the nine BIDs. Specifically, we use the total number of pairwise comparisons of genes in TCGA ($138,020,805$) to do the Bonferroni correction. Then we use the level 0.05.

\begin{table*}[!htbp]
\centering
\caption{Four contexts and the corresponding sample sizes.}
\label{data-contexts}
\begin{tabular}{@{}lrrrrc@{}}
\hline
Contexts& \multicolumn{1}{c}{All five subtypes}
& \multicolumn{1}{c}{LumA/LumB/Her2} & \multicolumn{1}{c}{Her2/LumB}
& \multicolumn{1}{c}{LumA}  \\
\hline
Sample Sizes   & 817 & 656  & 241  & 415  \\
\hline
\end{tabular}
\end{table*}

The BET detection results of TCGA breast cancer data for these four different contexts and five nonlinear BIDs are summarized in Section \ref{3.2}. More detailed descriptions are in Section \ref{3.3} through Section \ref{bio_pat}.

\subsection{Summary of BET Analysis}\label{3.2}

\begin{figure}[]
\centering
\includegraphics[width=0.8\textwidth,height=0.8\textheight,keepaspectratio]{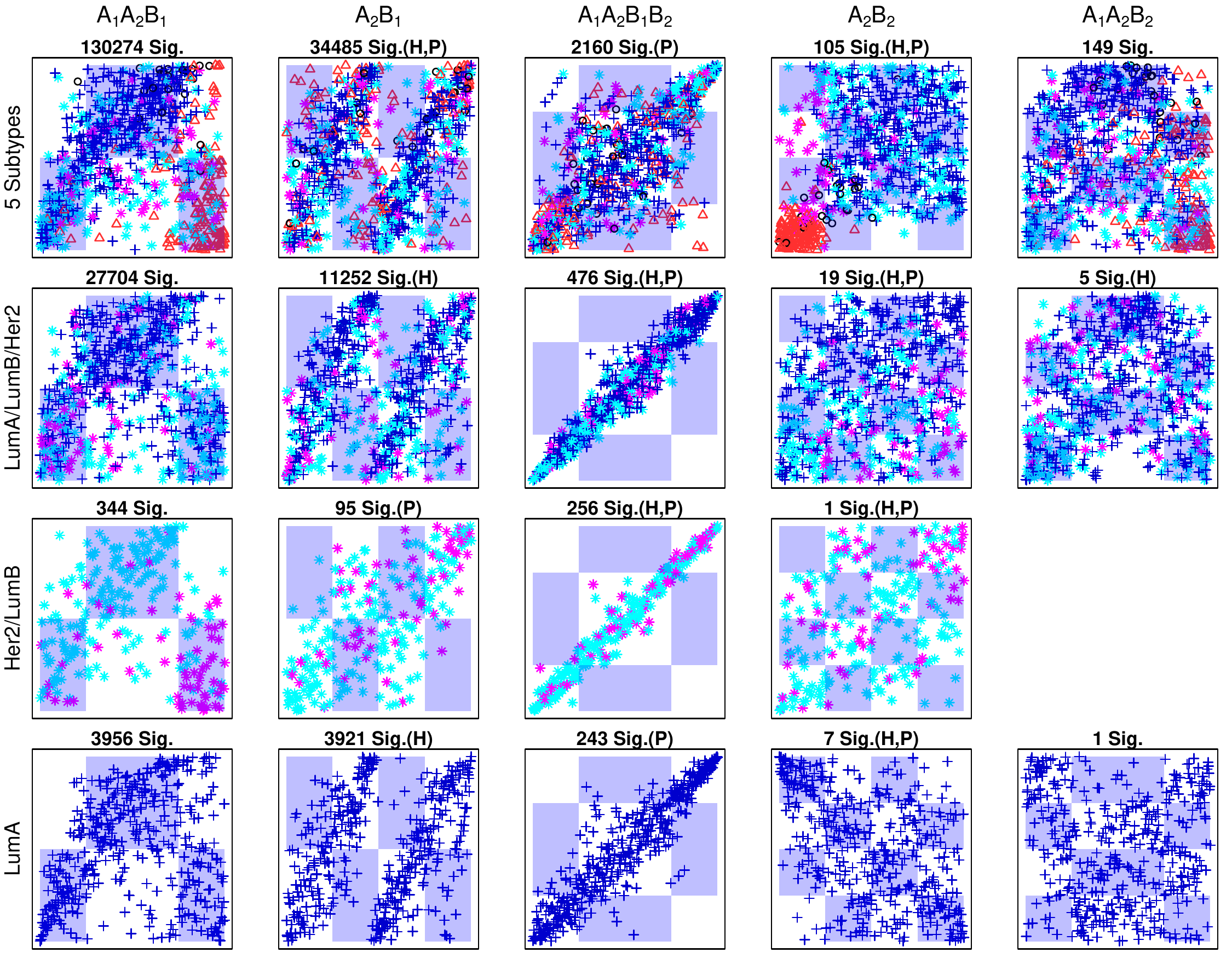}
\caption{BET diagnostic plots where the four contexts are the rows and the five nonlinear BIDs are the columns. The most significant pair is shown for each. The number of significant pairs for each BID is shown at the top. There is no significant pair for Her2/LumB at $A_1A_2B_2$. The symbols "H" and "P" at the top show that this most significant pair is Hoeffding's D and Pearson correlation respectively Bonferroni significant.}
\label{SummaryOfResults}
\end{figure}

As we discussed at the end of Section \ref{2.2}, there are three pairs of reflected nonlinear BIDs in all nine BIDs. Identifying such pairs results in five nonlinear BIDs, see the five columns of Figure \ref{SummaryOfResults}. From this point on, the \textit{Parabolic} BID shown in the first column refers to the union of this $A_1A_2B_1$ BID and its reflection $A_1B_1B_2$. In our TCGA analysis, this Parabolic BID frequently finds a Mixture data pattern. 
Similarly, the \textit{W} BID in the second column refers to the union of this $A_2B_1$ BID and its reflection $A_1B_2$, since the shape of this BID looks like the letter W. This W BID tends to find a particular bimodal pattern in our TCGA data. The rows of Figure \ref{SummaryOfResults} are the four contexts. The top gene pairs are shown for each. The number of significant pairs for each BID is displayed at the top of each panel. No pair is shown for the Her2/LumB context with the BID $A_1A_2B_2$ because there is no significant pair of genes.

Figure \ref{SummaryOfResults} shows the Parabolic BID in the first column, which contains the largest number of significant pairs for each context. In those cases, some obviously show mixtures of different subtype distributions, such as the red Basal subtype in the lower right region of the top Five Subtypes panel and the magenta Her2 cases in the lower right region of the Her2/LumB panel. The other relationships look strong but are not explained by subtypes, perhaps motivating additional genomic research. In particular, gene expression is indicative of many biological phenomena. Some are related to cancer subtypes, and some are not. A more detailed discussion of the gene pair in the upper left appears in Section \ref{3.3}; of the gene pair in the lower left in Section \ref{3.4}.

Some surprising bimodal patterns in column 2 are captured by the W BID, which seem not to be driven by the breast cancer subtype information. In Section \ref{bio_pat}, we discuss the gene pair on the top row of column 2. To investigate the potential biological relevance of this bimodal dependence, an independent data set is used and discussed in Section \ref{lack_rep}.

In column 3, some approximately linear pairs of genes are detected by the BID $A_1A_2B_1B_2$. In Section \ref{3.6}, we analyze the top pair in column 3 to discuss the connection between linearity and this nonlinear dependence pattern.

In column 4, as discussed in Section \ref{2.1}, we find an interesting biological separation in the Five Subtypes panel (top), which exists between the red Basal subtype and other breast cancer subtypes. Moreover, in the LumA context (bottom), this checkerboard pattern suggests that while many points are along the main diagonal, two small clusters lie off the main diagonal. This dependence pattern is not explained by the breast cancer subtypes and might be worth deeper biological investigations.

In column 5, the biological explanation of the most significant gene pairs shown in each context is not particularly clear. However, in Figure \ref{Figure-2}, the bottom-middle panel gives another example of this BID, which does contain an interesting biological pattern. We notice that the Basal points are clustered in the upper-left corner, which are well separated from other subtypes. Furthermore, the Luminal B (cyan star) subtype appears mostly at a cluster in the bottom center.

\begin{table*}[]
\centering
\caption{Comparison of BET, Hoeffding's D and Pearson Correlation over BET significant pairs of five BIDs in the Five Subtypes Context}
\label{compare}
\begin{tabular}{@{}lrrrrrc@{}}
\hline
BIDs& \multicolumn{1}{c}{$A_1A_2B_1$}
& \multicolumn{1}{c}{$A_2B_1$} & \multicolumn{1}{c}{$A_1A_2B_1B_2$}
& \multicolumn{1}{c}{$A_2B_2$}  & \multicolumn{1}{c}{$A_1A_2B_2$} & Total\\
\hline
Number of \\BET Significant Pairs   & 130,274  & 34,485  & 2,160  & 105 & 149 & 167,173 \\
\hline
Number of \\Pearson Correlation\\Significant Pairs   & 38,834  &  34,064    &  1,662  &  105  & 98 &  74,763 \\
\hline
Proportion of \\Pearson Correlation\\Significant Pairs   & 29.8\% & 98.8\%  & 76.9\% & 100.0\% & 65.8\% & 44.7\% \\
\hline
Number of \\Hoeffding's D \\Significant Pairs   & 41,292  & 18,228     & 1,071   & 56   & 58 & 60,705\\
\hline
Proportion of \\Hoeffding's D \\Significant Pairs   & 31.7\% & 52.9\%  & 49.6\% & 53.3\% & 38.9\% & 36.3\% \\
\hline
\end{tabular}
\end{table*}

Figure \ref{SummaryOfResults} also allows the comparison of BET with Hoeffding's D and Pearson correlation. In particular, the symbols "H" and "P" at the top of each panel indicate that the corresponding most significant gene pair is also eeHoeffding’s D independence/Pearson correlation testing Bonferroni significant across all $1.38\times 10^{8}$ possible pairs. From Figure \ref{SummaryOfResults}, we find that many interesting nonlinear pairs of genes discovered by BET are not Bonferroni significant when using Hoeffding's D or Pearson correlation, especially some biologically interesting Parabolic patterns (Column 1). Hoeffding’s D and Pearson tend to discover approximately linear dependence patterns. Based on this observation, an interesting question is how many of these nonlinear significant dependence pairs found by BET can not be discovered by conventional methods. This is studied in Table \ref{compare}. The pairs discovered by BET are assessed for significance by both classical linear Pearson correlation and the nonlinear Hoeffding's D. Recall that according to Table \ref{table1}, Hoeffding's D is about three times slower than BET. We compare these three methods over the significant pairs of genes of five nonlinear BIDs shown in Figure \ref{SummaryOfResults} in the Five Subtypes context. Specifically, for the Parabolic BID ($A_1A_2B_1$) in the Five Subtypes context, we apply Hoeffding's D over the $130,274$ BET significant pairs of genes and record the count of Pearson correlation and Hoeffding's D significant results after the Bonferroni adjustment for p-values across all possible pairs ($138,020,805$). Then we calculate the proportion of Pearson correlation and Hoeffding's D significant pairs in the BET significant pairs in each BID. Table \ref{compare} summarizes the number of BET significant pairs (first row), the number (second row) and the proportion (third row) of Pearson correlation significant pairs, the number (fourth row) and the proportion (fifth row) of Hoeffding's D significant pairs for each BID.

Note that less than $30\%$ of the pairs discovered by the Parabolic BID ($A_1A_2B_1$) were detected by Pearson correlation. For other patterns, the performances of Pearson correlation are better. On the other hand, about the comparison with Hoeffding's D, we notice that only $30\%-50\%$ BET significant pairs are detected by Hoeffding's D test in each of the five BIDs. Overall, $92,410$ and $106,468$ BET Significant Pairs are not discoverable by Pearson correlation and Hoeffding's D. This result indicates that BET is substantially more powerful against these particular types of nonlinear dependence.

\subsection{Mixture Pattern for Five Subtypes}\label{3.3}

\begin{figure}[!]
\centering
\includegraphics[width=0.5\textwidth,height=0.5\textheight,keepaspectratio]{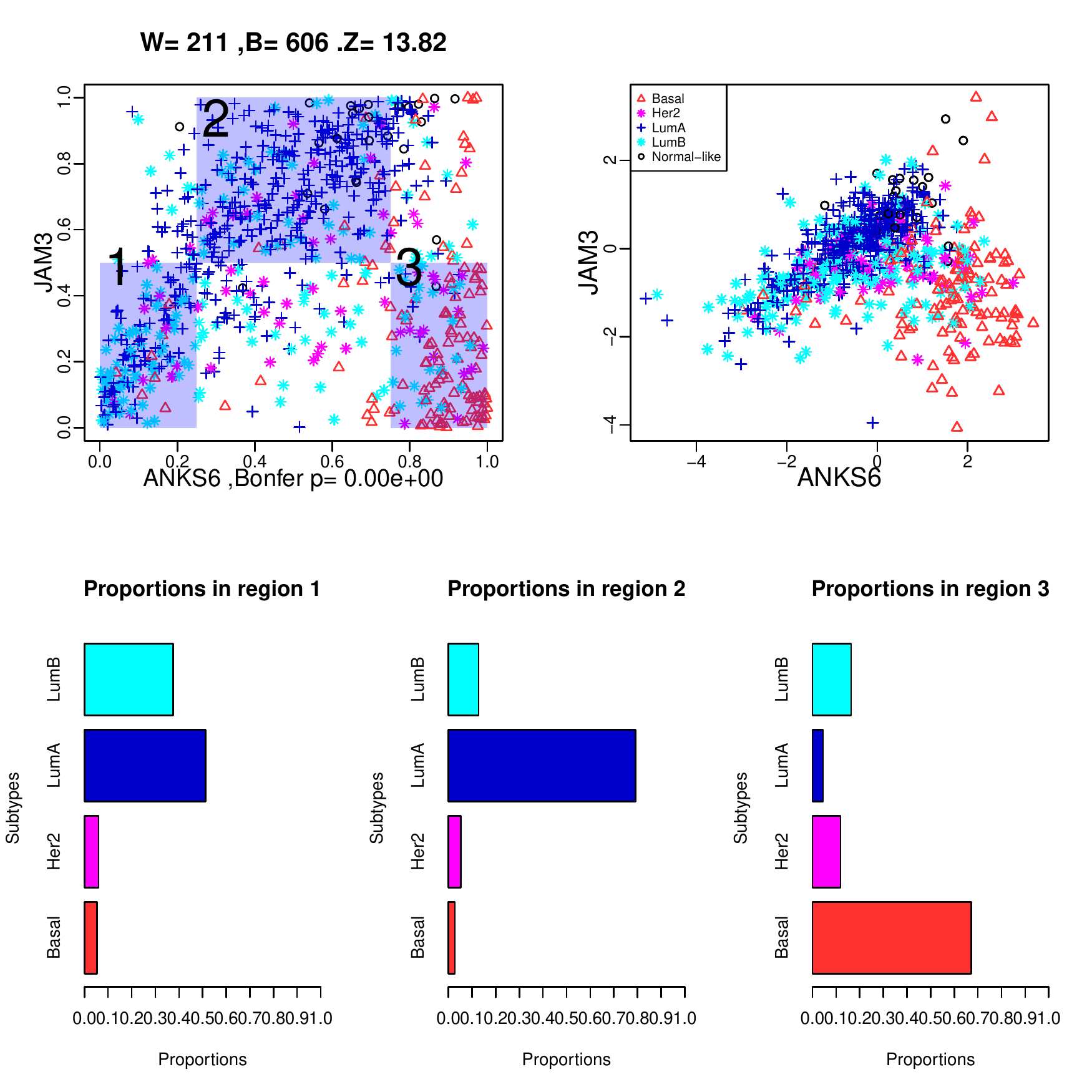}
\caption{Top-left: BET diagnosis plot for the pair of genes: ANKS6 and JAM3, which shows a strong surprising dependency with large z-statistic value and three blue rectangular regions; Top-right: the scatter plot of the same two genes in the normalized log count scale after pre-processing. Bottom: the proportions for each breast cancer subtype (except normal-like) in each blue rectangular region.}
\label{3_3_example}
\end{figure}

In this section, we discuss the pair of genes shown in the top panel of Figure \ref{3_3_example}: ANKS6 and JAM3, which has the most significant mixture dependence pattern. The dependence of these two genes is captured by the blue region of this BID. From the log-scale scatter plot (top-right) and BET diagnosis plot (top-left) in Figure \ref{3_3_example}, it is visually apparent that the Basal subtype group is separated from the other subtypes. In particular, the non-Basal cases look like a classical bivariate Gaussian distribution with a \textit{positive correlation} between ANKS6 and JAM3 (larger values of ANKS6 lead to more expression of JAM3). However, the Basal cases behave very differently: larger ANKS6 goes along with smaller JAM3, indicating a \textit{negative correlation}. Ignoring this important difference in direction of correlation can have a serious impact on gene network analysis. ANKS6 in this example is the main driver of the separation between Basal and the others. 

Here we revisit the reflection issue from the end of Section \ref{2.2}. As discussed above, the BIDs $A_1A_2B_1$ and $A_1B_1B_2$ are identical if we switch the genes on the x-axis and the y-axis. Since the pair of genes in this example (ANKS6 and JAM3) is detected by the $A_1A_2B_1$ BID, reversing the ordering of these two genes will give the reflected pattern ($A_1B_1B_2$). This example explains why only five nonlinear BIDs are considered here.

Further insight comes from splitting the blue region into three rectangular regions (see the numerical labels in the top-left panel) and calculating the respective proportions of the four breast cancer subtypes (ignoring the Normal-like) in each blue rectangle (bottom panels). These proportions reveal how the subtypes drive this relationship.

In the proportion bar plots, LumA and LumB subtypes have high proportions in Regions 1 and 2. Those two subgroups account for $88.7\%$ and $92\%$ in the two regions separately, where the Basal subgroup only accounts for $5.4\%$ and $2.8\%$. However, in Region 3, the proportion of Basal reaches $67\%$, and the total proportion of LumA and LumB is $21\%$. This observation indicates that the positive correlation of the LumA and LumB domains are in Regions 1 and 2, and the differing Basal correlation is in Region 3. The Pearson Chi-square test of independence is used to confirm this observation, for the counts of points in Table \ref{3_3_Observed_table_1}.

\begin{table*}[!]
\centering
\caption{Observed counts for four breast cancer subtypes in three blue regions.}
\label{3_3_Observed_table_1}
\begin{tabular}{@{}lrrrrc@{}}
\hline
Subtypes & region 1 & region 2 & region 3\\
\hline
Basal   & 9    & 8    & 90 \\
LumA   & 86 & 229 & 6 \\
LumB   & 63 & 37 & 22 \\
Her2    & 10 & 15     & 16\\
\hline
\end{tabular}
\end{table*}  

The p-value of this Chisq test is smaller than $2.2\times 10^{-16}$ (i.e., smaller than floating-point round-off error), and it shows a strong significance that these four subtypes are not homogeneously distributed in the blue regions. To more directly validate the separation between Basal and the others, all others are combined into a single group and the Chi-square test gives another small p-value less than round-off error. This result confirms the observation that this dependence pattern captured by the blue regions is very strongly significant and is influenced by the mixture of Basal and other subtype distributions.

\begin{figure}[]
\centering
\includegraphics[width=0.38\textwidth,height=0.38\textheight,keepaspectratio]{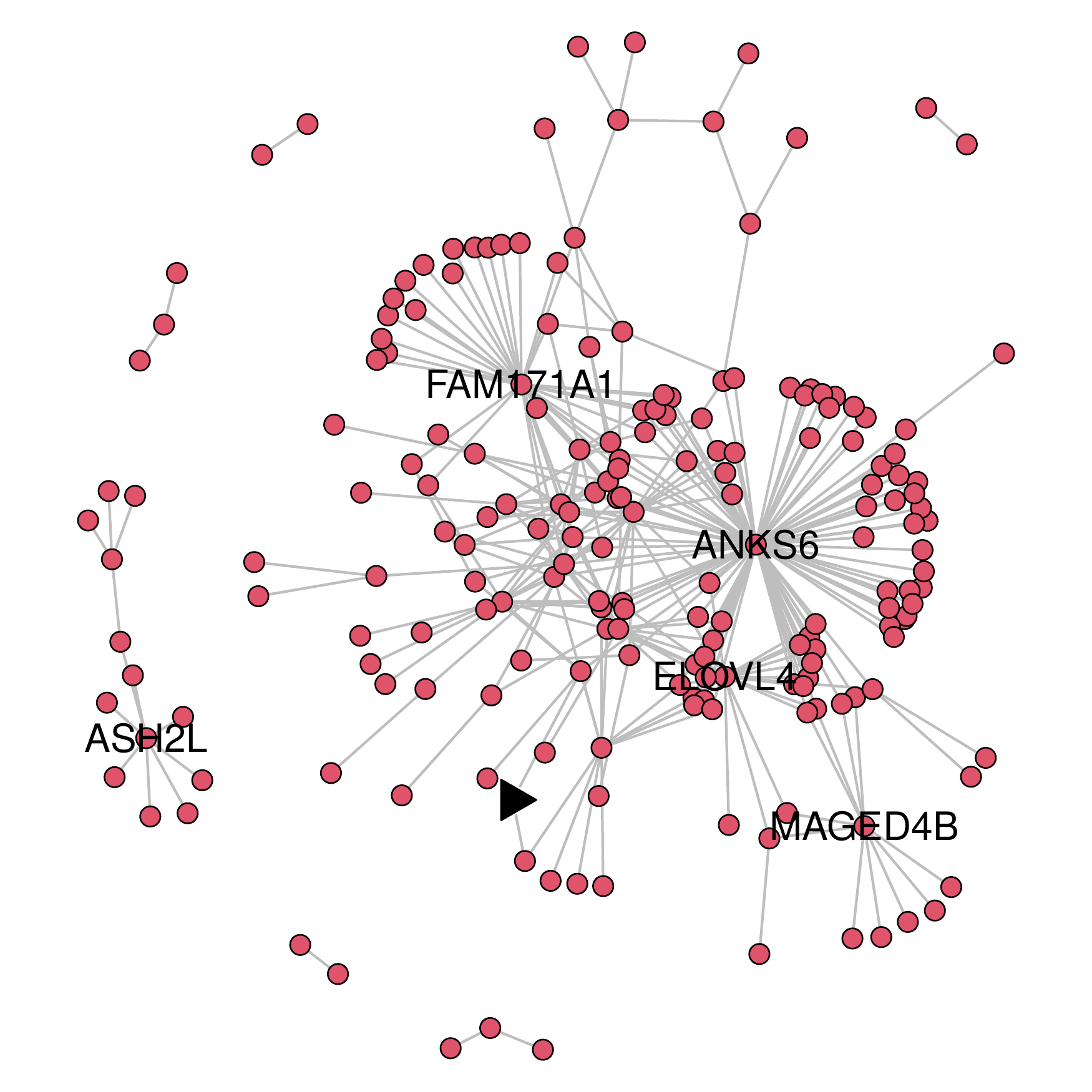}
\caption{Network connection plot of 200 most significant genes (nodes) with 311 edges for the BID $A_1B_1B_2$. Each node represents one gene and each edge represents a significant dependence between those genes. The large community illustrated there are many genes having significant mixture dependence pattern with the center gene, such as ANKS6. The point represented by the black triangle is the gene ZDHHC2 featured in Figure \ref{3_4_network}.}
\label{3_1_network}
\end{figure}

The relationships between pairs of genes with respect to this same BID $A_1B_1B_2$ is shown by a network connection plot in Figure \ref{3_1_network}. Nodes in Figure \ref{3_1_network} represent genes. This particular pattern of nonlinear BET dependence is highlighted by edges, which show Bonferroni statistical significance between genes. Furthermore, genes are ranked by their maximum BET z-scores. As noted in \cite{zhang2019BET}, z-scores, reflecting the number of standard deviations above the mean, are more interpretable when p-values are extremely small. As shown in Figure \ref{SummaryOfResults}, there are $130,274$ significant pairs of genes for this BID and its reflection. To avoid a too cluttered network graphic, only the top 200 genes are shown in Figure \ref{3_1_network}. These two hundred genes have 311 significant dependence edges for this Parabolic BID. Genes at the center of some visually important communities are labeled. Each community is a set of genes that shows this relationship with the center gene. Notice there are a number of gene communities representing different biological dependencies that are significant with respect to this BID. Figure \ref{3_3_example} suggests that much of this nonlinear dependence may be due to the Basal subtype which is well known to be quite different from the others. However, there can be other causes of this pattern. For example, Figure \ref{3_4_network} shows that the gene ZDHHC2 has such dependence even when the Basal subtype is left out of the analysis. This gene appears in Figure \ref{3_1_network} as the point represented by the black triangle.

Good insights into any of these gene communities and their functions come from finding where they appear among published gene signatures, such as those shown in The Molecular Signatures Database (MSigDB), a collection of annotated gene sets for use with Gene Set Enrichment Analysis \citep{GSEASubramanian2005,MSigDBLiberzon2011,MSigDBHallmarkCollectionLIBERZON2015417}. For example, we performed gene set enrichment analysis on the labeled communities. The gene set in the largest (ANKS6) community is strongly associated with stromal or immune features. This is consistent with the previous finding that basal-like breast cancer has increased immune signature expression \citep{Iglesia2016immunecell}. The gene analysis of the other label communities did not give such good biological interpretations.

\subsection{Mixture Pattern for Only Luminal A subtype}\label{3.4}

\begin{figure}[h]
\centering
{%
\resizebox*{4cm}{!}{\includegraphics[width=1\textwidth,height=1\textheight,keepaspectratio]{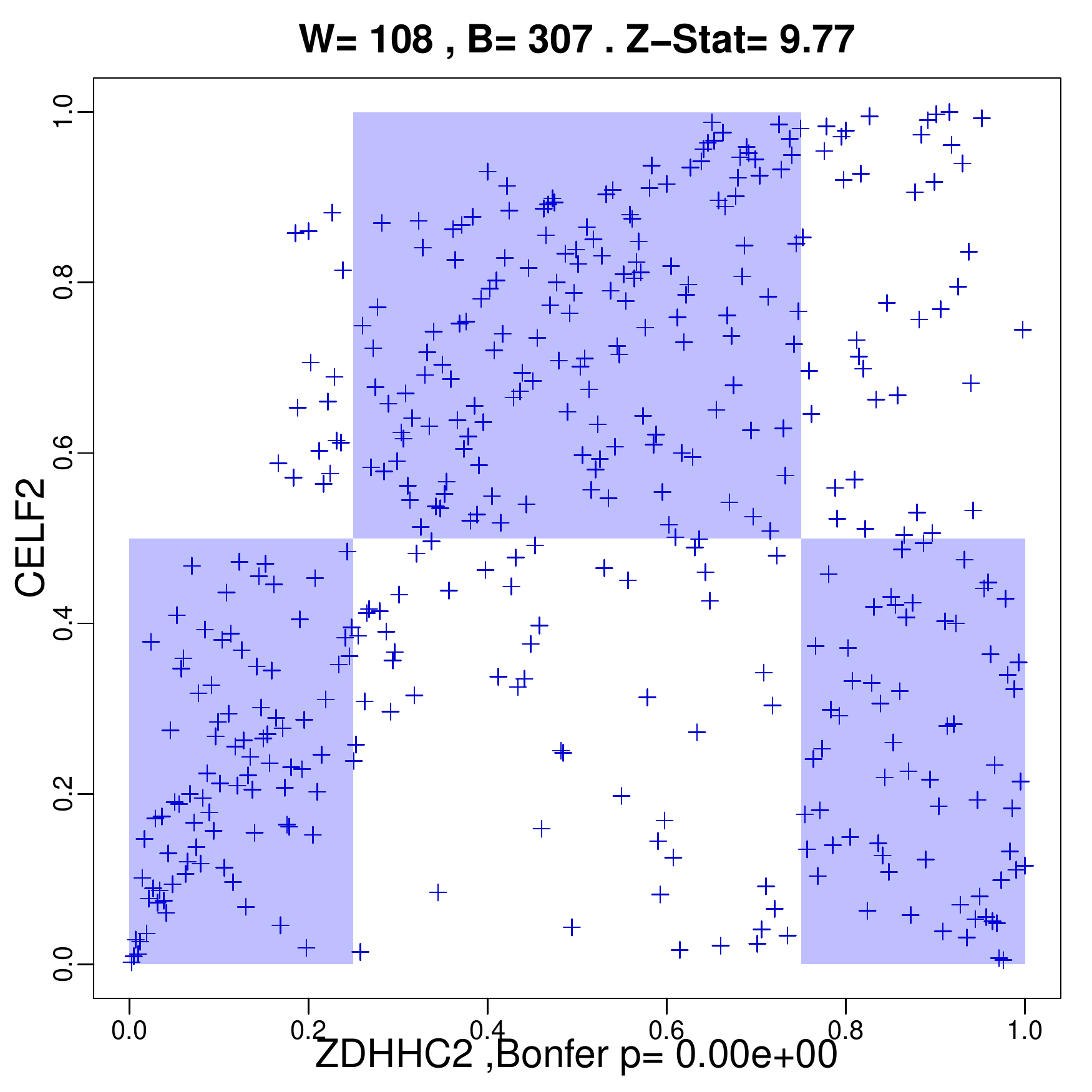}}}\hspace{5pt}
{%
\resizebox*{4cm}{!}{\includegraphics[width=1\textwidth,height=1\textheight,keepaspectratio]{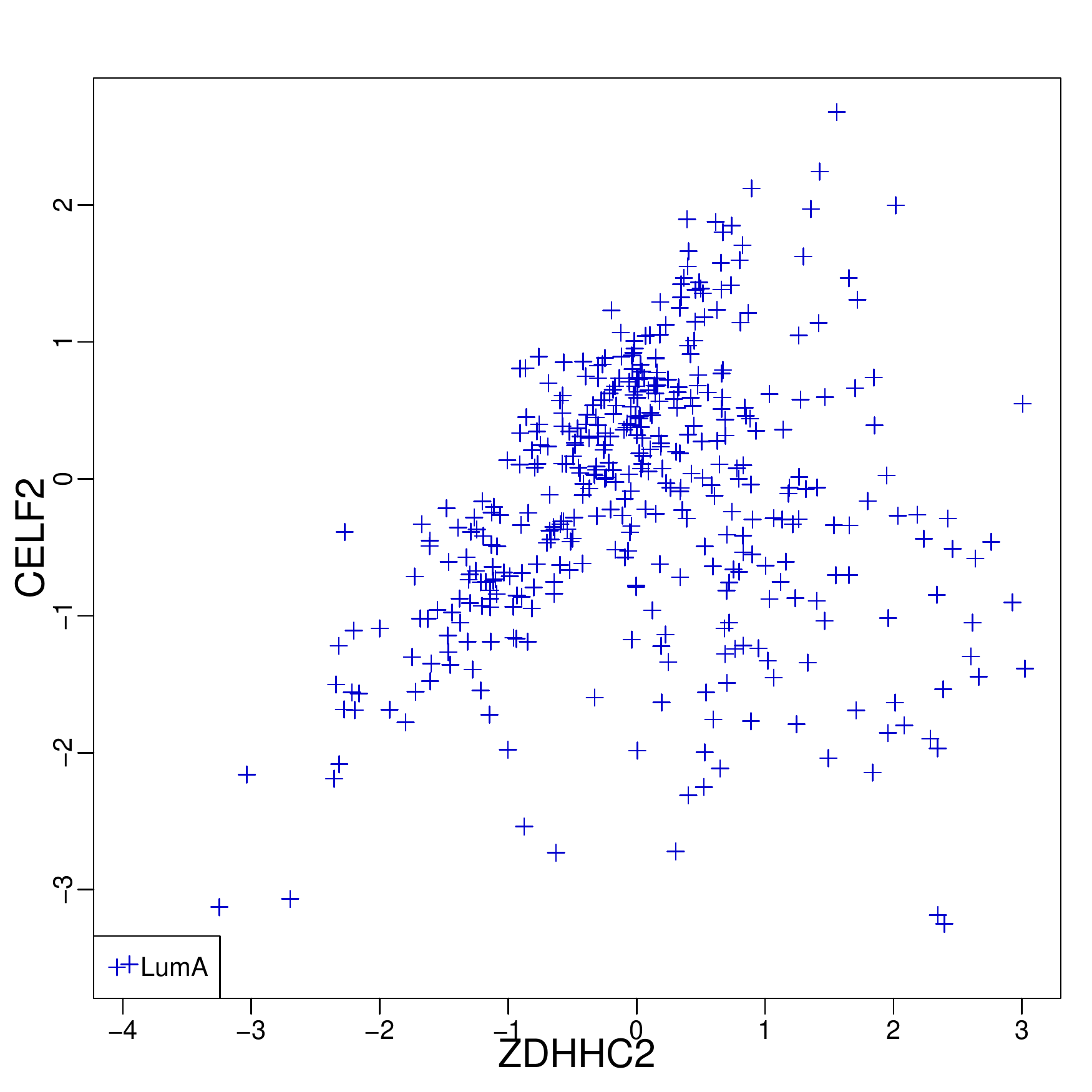}}}
\caption{Left: BET diagnosis plot for a pair of genes which shows the mixture dependence within only the Luminal A subgroup; Right: The scatter plot of the same two genes in the normalized count scale. A point cloud with a strong positive correlation is on the left of the entire group and the remaining cases form a more diffuse cluster on the bottom-right.} 
\label{3_4_example}
\end{figure}

While subtypes have played an important role in the diagnosis and treatment of breast cancer, the heterogeneity of the disease motivates deeper investigation within subtypes. Here we focus only on the Luminal A breast cancer subtype observations. Figure \ref{3_4_example} shows an additional interesting mixture dependence pattern in both the BET diagnosis (left panel) and log-scale scatter (right panel) plots. The right panel contains a positively correlated Gaussian point cloud on the left. There is a more diffuse cluster towards the lower right. This seems to indicate a mixture behavior. In particular, the gene ZDHHC2 bifurcates the data into a cluster where it is strongly positively correlated with CELF2, and another cluster where large values of ZDHHC2 correspond to small values of CELF2. Hence this pair of genes highlights potentially interesting subgroups, which merits a deeper investigation.

\begin{figure}[]
\centering
\includegraphics[width=0.37\textwidth,height=0.37\textheight,keepaspectratio]{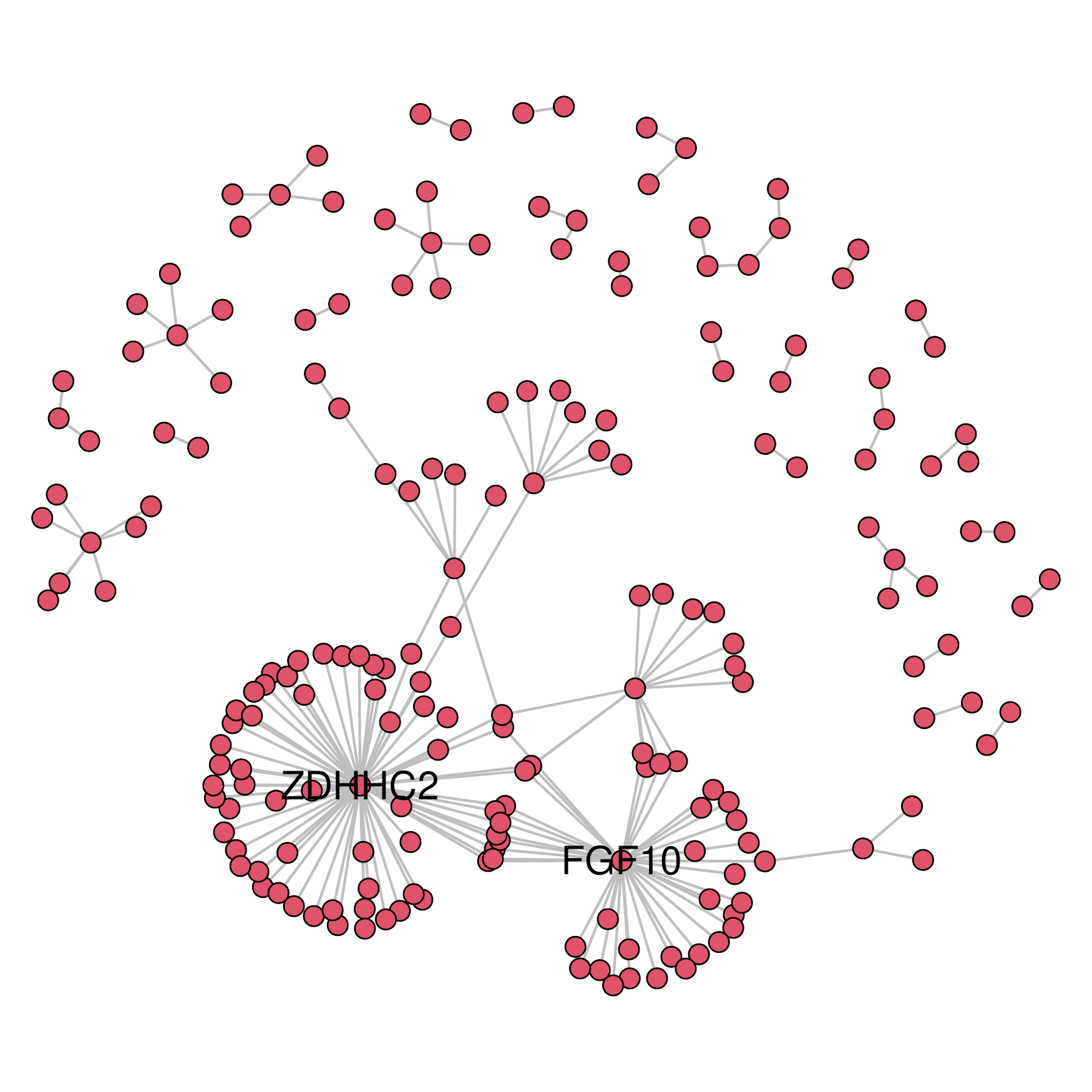}
\caption{Connection plot of 200 most significant genes (nodes) with 190 edges for the Parabolic BID in the context of only Luminal A. ZDHHC2 and FGF10 are central genes in two large communities. The ZDHHC2 gene community has overlaps with some luminal gene sets in The Molecular Signatures Database (MSigDB).}
\label{3_4_network}
\end{figure}

The connection plot in Figure \ref{3_4_network} shows which genes have many significant pairs within the 200 most significant genes in the LumA only context with the Parabolic BID. There are 190 significant dependence edges. This can be less than 200 because there are many pairs that only connect with each other. ZDHHC2 and FGF10 are two central genes having large communities, which motivate a deeper investigation. Checking carefully the individual plots reveals in all of these pairs, ZDHHC2 and FGF10 play the bifurcating role shown in Figure \ref{3_4_example} in the dependence with each of these other genes. To further illustrate this bifurcation property, we show more examples of pairs about ZDHHC2 for this pattern in Supplement "More Examples for Gene ZDHHC2 in the LumA Only Context for $A_1B_1B_2$". 

To investigate the corresponding gene function, we use gene set enrichment analysis again to compute overlaps between these communities and gene sets from MSigDB. This ZDHHC2 community has overlaps with some gene sets related to breast cancer, and in particular the luminal subtype, such as CHARAFE\_BREAST\_CANCER\_LUMINAL\_VS\_MESENCHY -MAL\_DN and CHARAFE\_BREAST\_CANCER\_LUMINAL\_VS\_BASAL\_DN \citep{BreastCancerLumVsMesen_charafejauffret:hal-01431970}. 
This confirms that ZDHHC2 and its community are an important player at luminal breast cancer and could motivate a deeper investigation into the role played by the ZDHHC2 community.
On the other hand, a similar investigation of FGF10 doesn't show the connection with research to date on luminal breast cancer, again possibly motivating further biological work.

\subsection{Connection between Linear and Nonlinear Patterns}\label{3.6}

\begin{figure}[]
\centering
{%
\resizebox*{4cm}{!}{\includegraphics[width=0.5\textwidth,height=0.5\textheight,keepaspectratio]{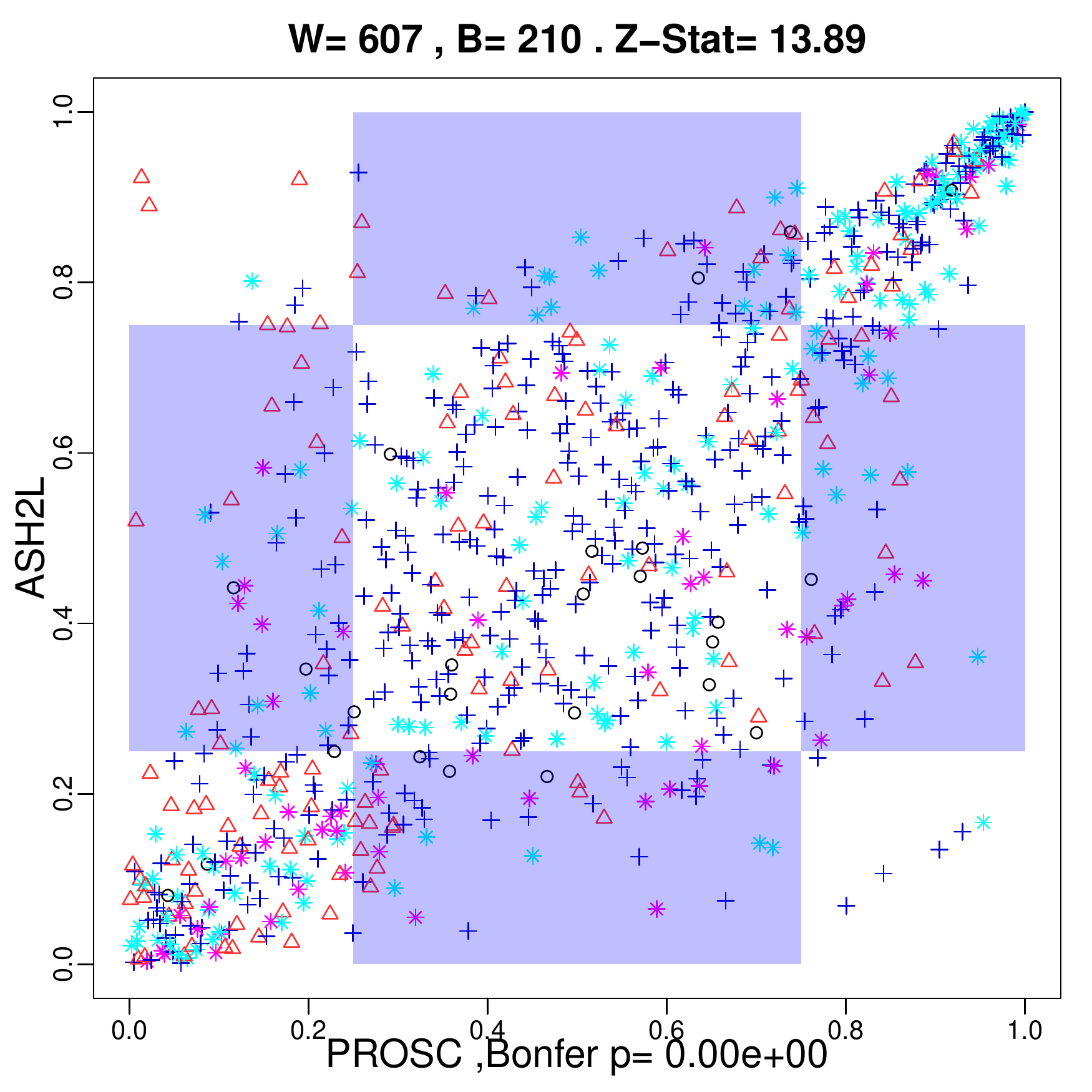}}}\hspace{5pt}
{%
\resizebox*{4cm}{!}{\includegraphics[width=0.5\textwidth,height=0.5\textheight,keepaspectratio]{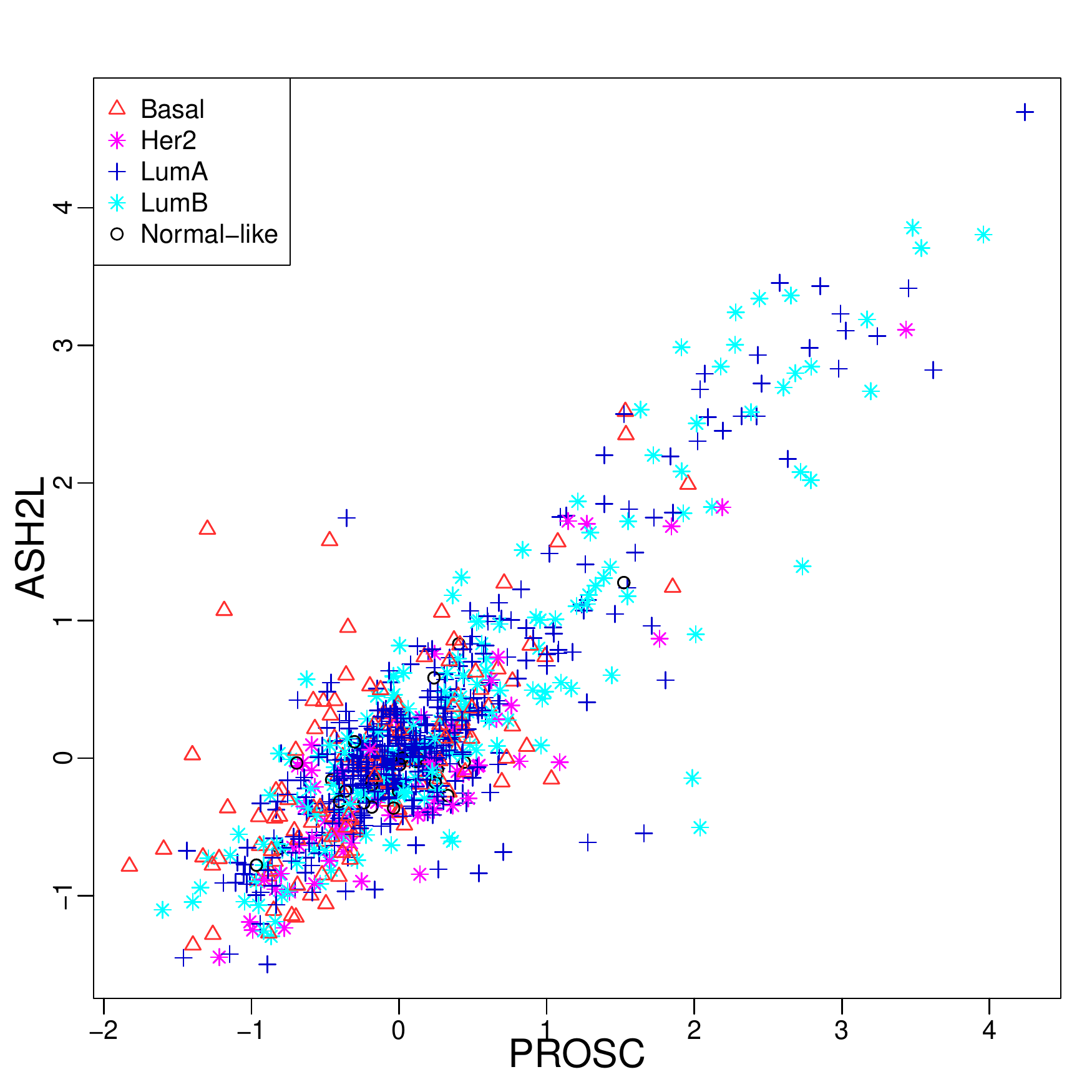}}}\hspace{5pt}
{%
\resizebox*{4cm}{!}{\includegraphics[width=0.5\textwidth,height=0.5\textheight,keepaspectratio]{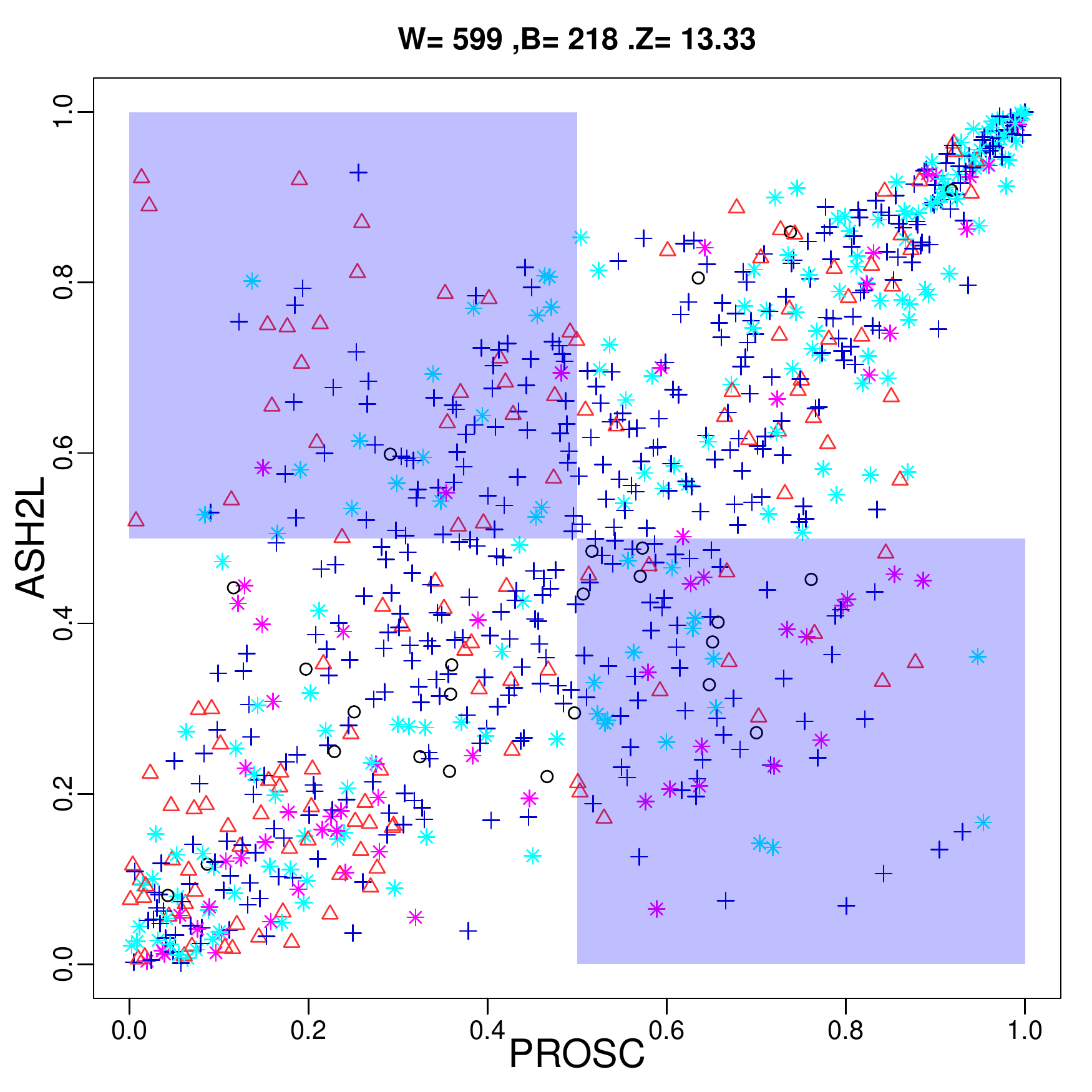}}}
\caption{Left: The most significant BID diagnosis plot for a pair of genes: PROSC and ASH2L; Middle: The scatter plot of the same two genes in the normalized count scale; Right: The corresponding linear BID $A_1B_1$ for the same two genes. The value of each BET statistic for the linear BID is slightly smaller than that for the most significant BID.} \label{3_6_example}
\end{figure}

In column three of the summary plot Figure \ref{SummaryOfResults}, 
some approximately linear pairs of genes with a second-order structure 
are detected by the BID $A_1A_2B_1B_2$. The left panel of Figure \ref{3_6_example} gives the BID diagnosis of the top row pair of genes in Column 3 (PROSC and ASH2L) with the largest symmetry statistic value and z-score in all nine BIDs. As discussed in Section \ref{2.1}, this BET diagnosis plot shows that this pair is not bivariate normal with concentrations in both ends of the diagonal. A deeper investigation of the structure of this pair of genes is from the scatter plot in the middle panel of Figure \ref{3_6_example}. It reflects an approximate linear relationship with strong skewness along the major axis. To analyze the connection between this pair and the linear dependence BID $A_1B_1$, we show the linear BID in the right panel. The corresponding counts of white and blue regions for BID $A_1A_2B_1B_2$ in the left panel are $607$ and $210$, so that the symmetry statistic S and the z-score are $397$ and $13.89$; the corresponding counts for the linear BID in the right panel are $599$ and $218$, so that the symmetry statistic S and the z-score are rather close, but slightly smaller values of $381$ and $13.33$. These numbers reflect the unusual pattern of greater variation in the middle of the distribution, with relatively less variation from the diagonal for the rest of it. This suggests a different type of mixture model which may merit deeper investigation.

\subsection{Bimodal Pattern}\label{bio_pat}
The second column of Figure \ref{SummaryOfResults} shows a perhaps surprising bimodal dependence pattern that is shared by many pairs of genes.

\begin{figure}[H]
\centering
{%
\resizebox*{4cm}{!}{\includegraphics[width=1\textwidth,height=1\textheight,keepaspectratio]{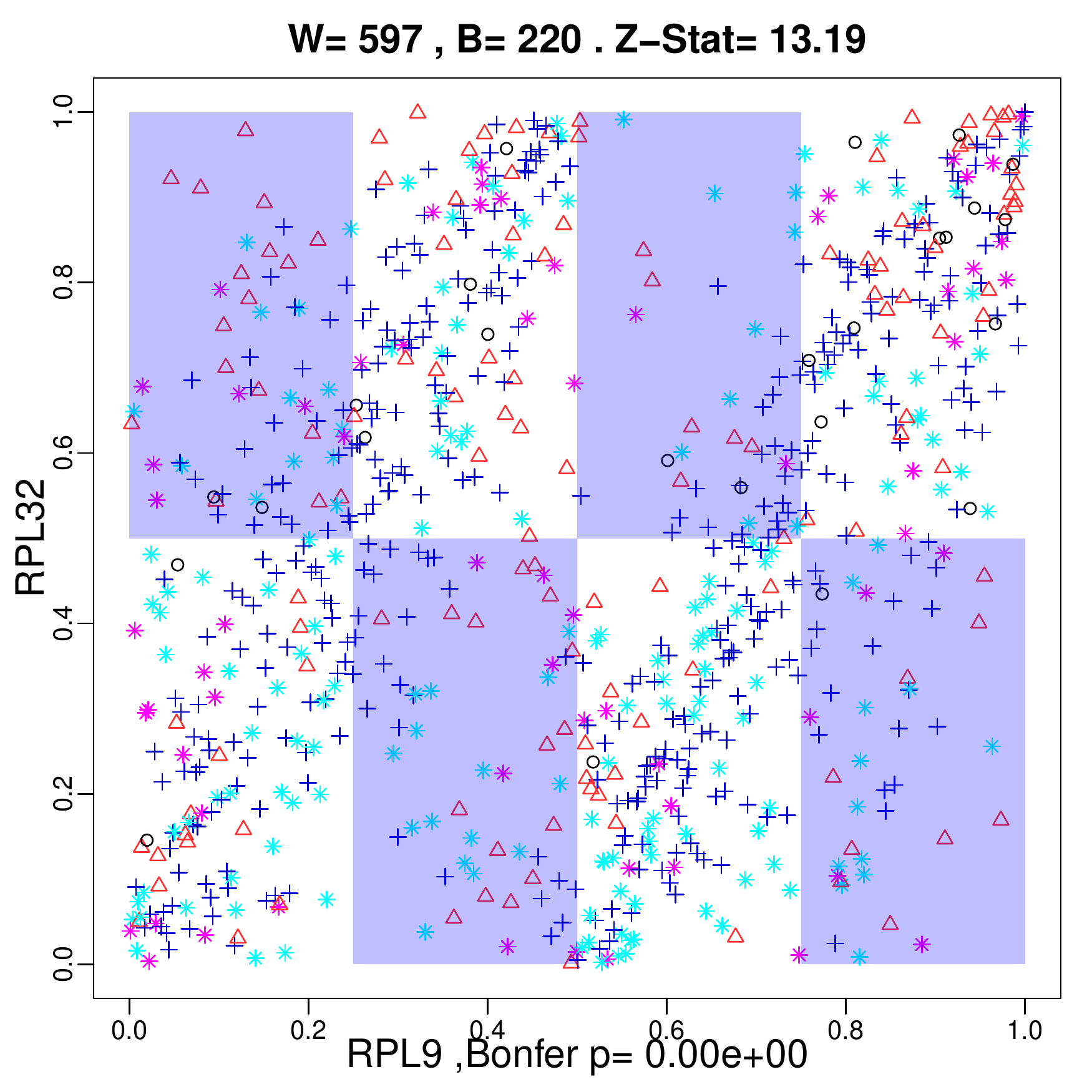}}}\hspace{5pt}
{%
\resizebox*{4cm}{!}{\includegraphics[width=1\textwidth,height=1\textheight,keepaspectratio]{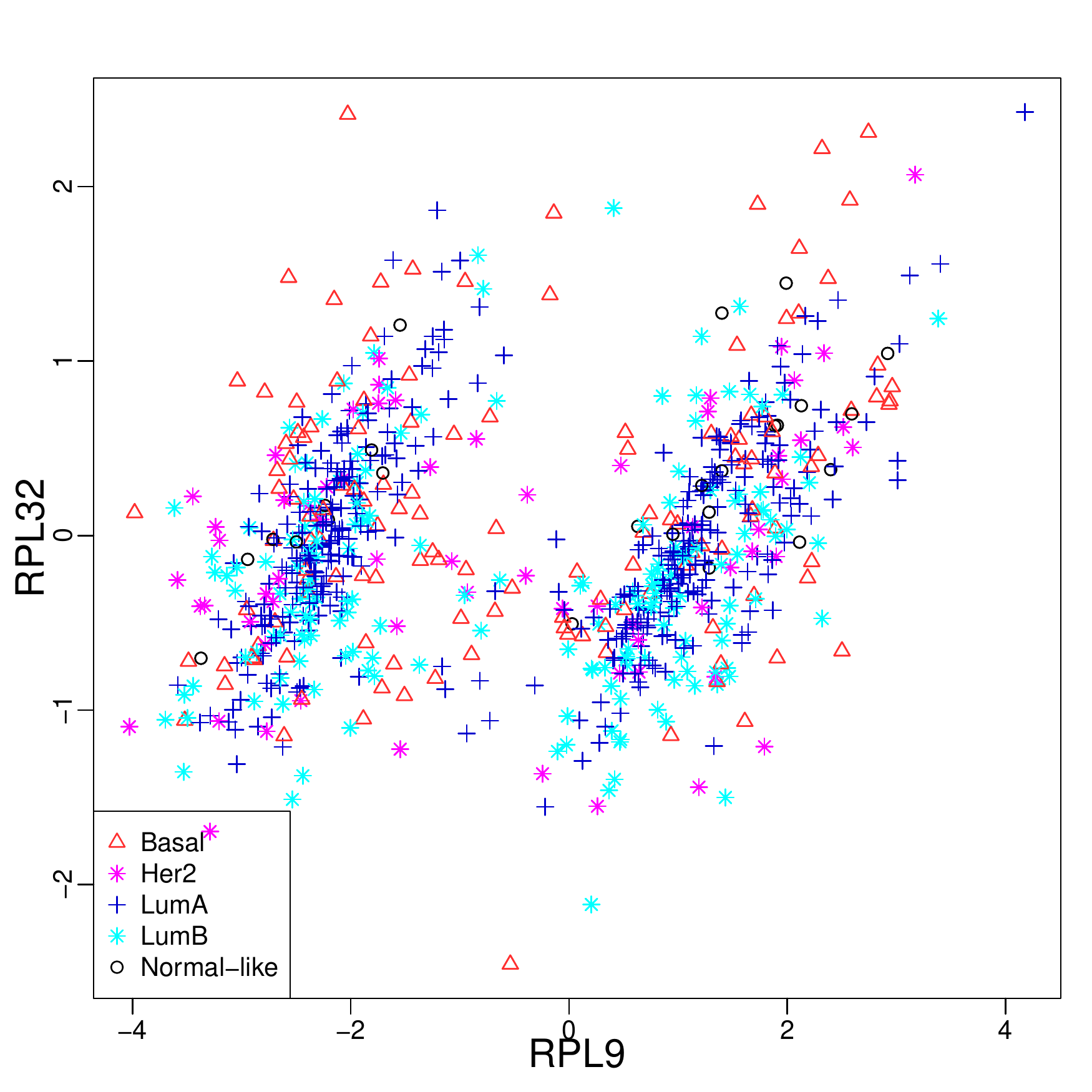}}}
\caption{Left: BET diagnosis plot for a pair of genes that shows a bimodal dependence pattern; Right: The scatter plot of the same two genes in the normalized count scale. This nonlinear relationship seems to be driven by the bimodal distribution of gene RPL9.} \label{3_5_example}
\end{figure}

To more deeply investigate this bimodal dependence, we take the top pair in the context of Five Subtypes as an example. Figure \ref{3_5_example} shows the BET diagnosis (left panel) and log-scale scatter (right panel) plots of the most significant pair of this pattern in the context of Five Subtypes: RPL9 and RPL32. RPL9 in this pair separates the group into two positively correlated Gaussian clusters, suggesting this surprising bimodal dependence pattern perhaps is related to the gene RPL9. To investigate whether this is an important biological phenomenon or an artifact of some particular preprocessing steps in the Lobular Freeze TCGA data, we consider an additional completely separate data set in Section \ref{lack_rep}.

\section{Biological Reproducibility of TCGA Results}\label{4}
An essential issue with exploratory data analyses, as shown in Section \ref{3} is their reproducibility. To investigate this for the interesting results shown in Sections \ref{3.3} and \ref{bio_pat}, we consider an independent genomic data set: the Sweden Cancerome Analysis Network-Breast (SCAN-B) \citep{SCAN-B}. This data set came from the NCBI Gene Expression Omnibus (GSE96058). The Data set was preprocessed as described in \citep{Saal2015}. We use a subset of the gene expression data set which contains 2969 samples with full clinical data and 30865 genes. There are $15,197$ genes existing in both the SCAN-B and TCGA data sets. We only consider these common genes during this validation process. There was no further processing step in the SCAN-B set for our reproducibility analysis. First, in Section \ref{4.1}, we study the reproducibility of the Mixture patterns in the contexts of the Five Subtypes, as shown in Section \ref{3.3}. Then in Section \ref{lack_rep}, we find that the bimodal distribution of the gene RPL9 in Section \ref{bio_pat} is not observed in SCAN-B. This discrepancy is explained in Supplement D.

\subsection{Biological Reproducibility of the Mixture Pattern}\label{4.1}

\begin{figure}[]
\centering
\includegraphics[width=0.75\textwidth,height=0.75\textheight,keepaspectratio]{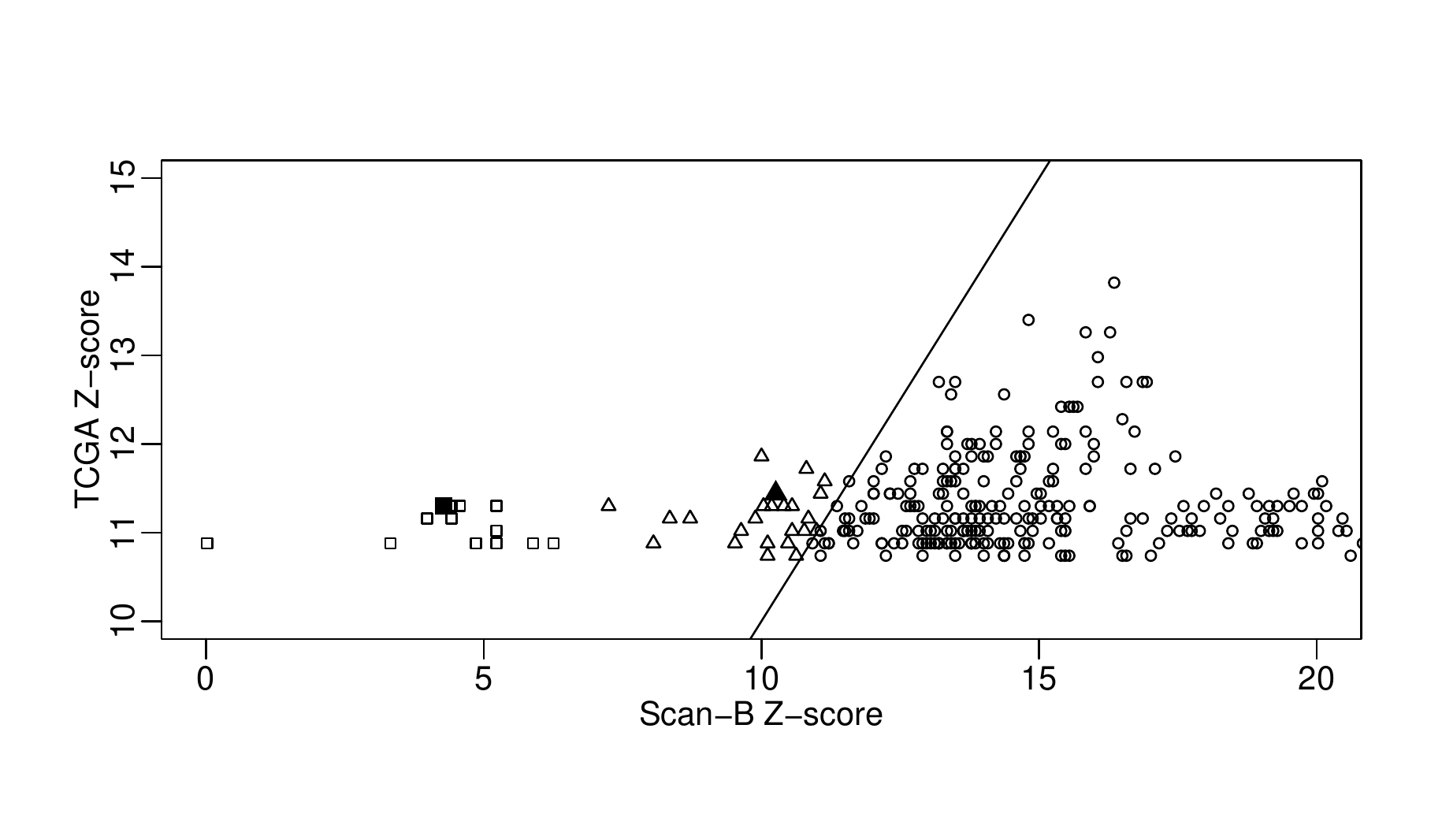}
\caption{Scatter plot comparing the Parabolic BID z-scores between the SCAN-B and TCGA data sets for the significant Mixture pairs within the top 200 genes in the Five Subtypes context of TCGA. The dark square and dark triangle points are illustrated in Figures \ref{4_2_example_2} and \ref{4_2_example_1}.}
\label{4_2_comparison}
\end{figure}

In Section \ref{3.3}, we find an interesting mixture pattern detected by the Parabolic BID in the context of the Five Subtypes. To investigate whether this mixture pattern is reproducible in the SCAN-B data, we chose the most significant 200 genes for this context in TCGA results, as shown in the network connection plot Figure \ref{3_1_network}. To understand the relationship between pairs over the two data sets, we rerun BET for these pairs in the SCAN-B data set and record the corresponding z-scores for the Parabolic BID. Thus, we compare the significance for only the mixture pattern in these two data sets, as shown in Figure \ref{4_2_comparison}.

Within these top 200 genes, 7 genes do not exist in the SCAN-B. Considering only the remaining 193 genes resulted in 298 significant (in TCGA) mixture pairs. The corresponding SCAN-B significance for each pair is compared in Figure \ref{4_2_comparison}. In particular, each point is one mixture pair whose TCGA z-score is shown on the vertical axis and SCAN-B z-score is shown on the horizontal axis.

Because the sample size is much larger for SCAN-B, stronger significance is expected for most pairs. This is highlighted using the dark line $y = x$ showing which pairs are equal. As expected, most SCAN-B z-scores are larger, reflected as circle points to the right of the line $y = x$.

\begin{figure}[]
\centering
{%
\resizebox*{4cm}{!}{\includegraphics[width=0.5\textwidth,height=0.5\textheight,keepaspectratio]{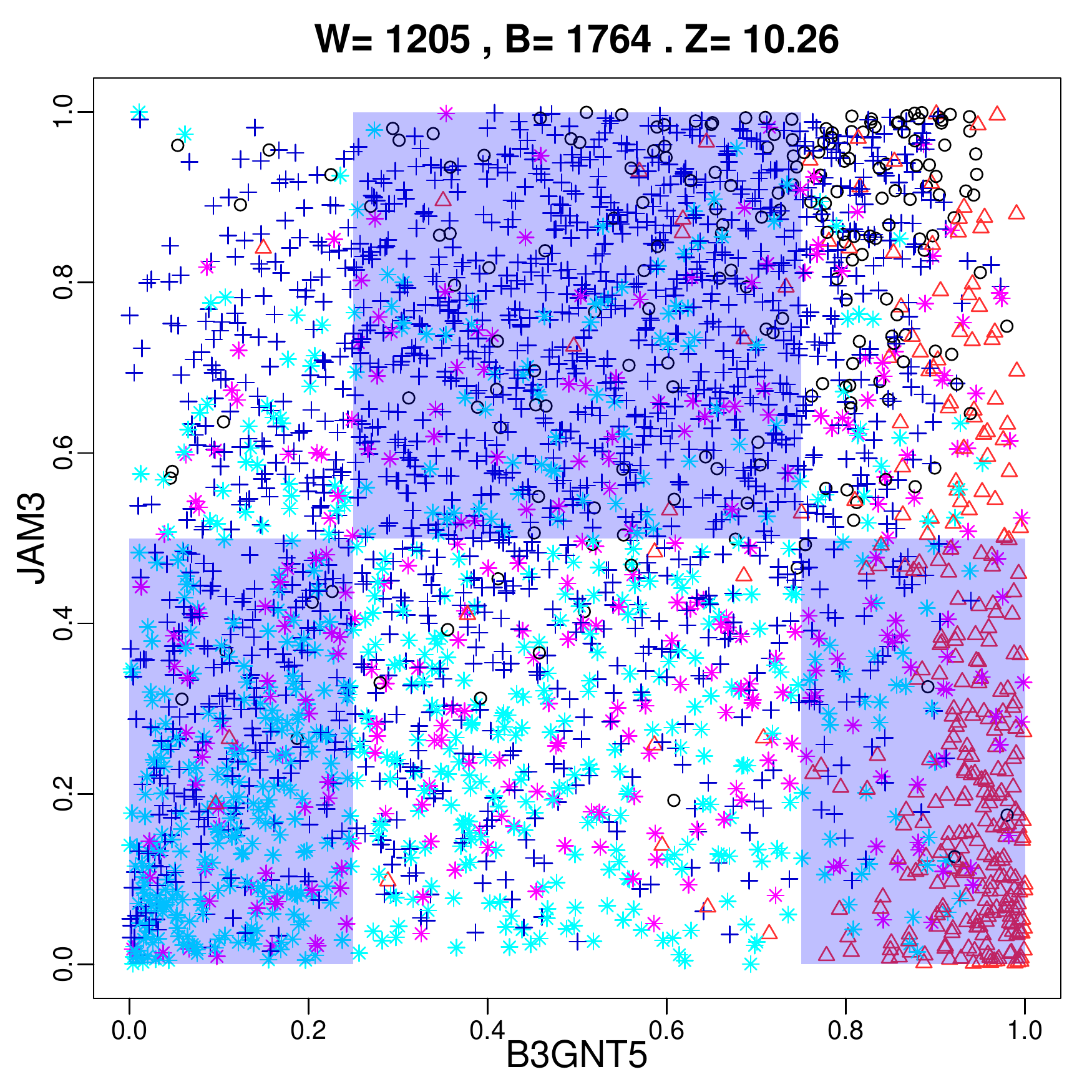}}}\hspace{5pt}
{%
\resizebox*{4cm}{!}{\includegraphics[width=0.5\textwidth,height=0.5\textheight,keepaspectratio]{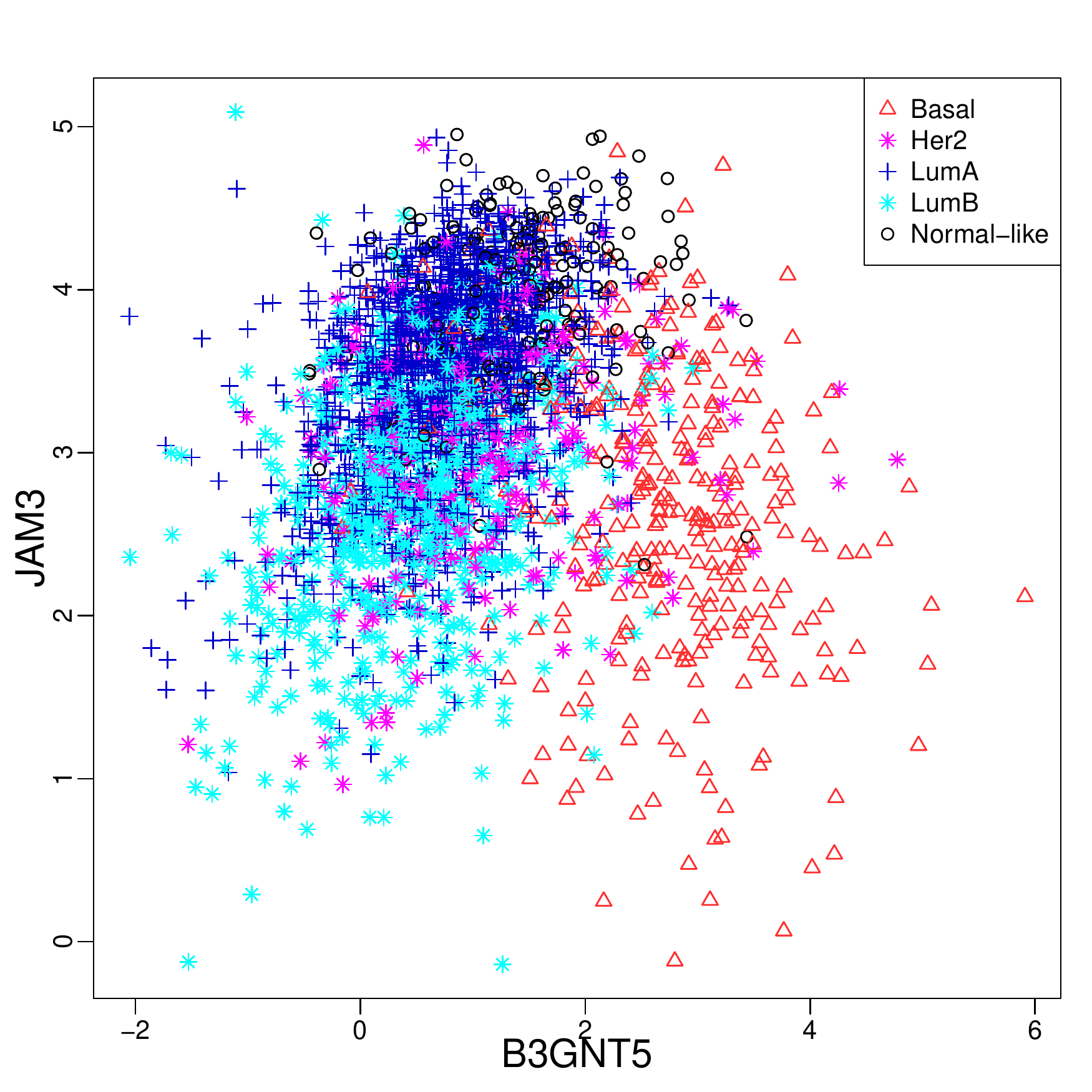}}}\hspace{5pt}
{%
\resizebox*{4cm}{!}{\includegraphics[width=0.5\textwidth,height=0.5\textheight,keepaspectratio]{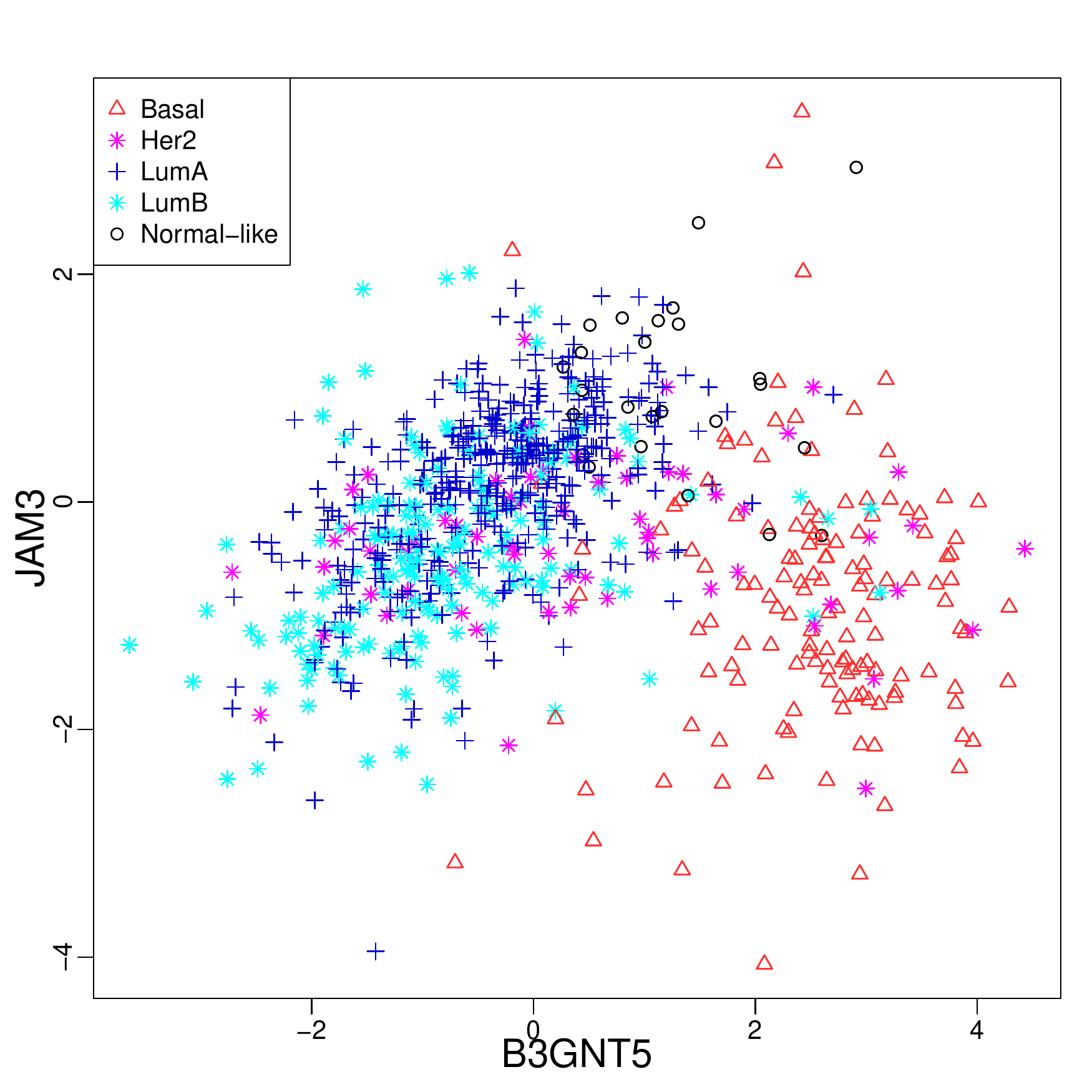}}}
\caption{Left: BET diagnosis plot for a pair of genes in the SCAN-B data set that is shown as a dark triangle in Figure \ref{4_2_comparison}; Middle: The scatter plot of the same two genes in the original SCAN-B scale; Left: The scatter plot of the same two genes in the normalized TCGA count scale. This example represents gene pairs which tend to be less significant in the SCAN-B data set but still have the mixture pattern.} \label{4_2_example_2}
\end{figure}

Points to the left of the line $y = x$ seem to fall into two different types. For the triangle points, the TCGA z-score is not much bigger than the SCAN-B z-score, suggesting this could be just random variation. This is more carefully investigated in Figure \ref{4_2_example_2}.
The square ones are investigated in Figure \ref{4_2_example_1}. These are gene pairs with substantial missing values in the SCAN-B version of the data.

Typical behavior of the triangle points is studied in Figure \ref{4_2_example_2} by showing the pair represented as a dark triangle in Figure \ref{4_2_comparison} . The BET diagnosis (left panel) and the SCAN-B scatter (middle panel) plots definitely show the same behavior as in Figure \ref{3_3_example}, which is a clear separation between Basal and the other subtypes.
However, the separation of the Basal is more distinct in TCGA, as shown in the TCGA scatter plot (right panel) of the same two genes, which is consistent with the more significant TCGA z-scores. We observed similar behavior for each of the pairs represented as triangles in Figure \ref{4_2_comparison}.

Figure \ref{4_2_example_1} shows the pair highlighted with the dark square in Figure \ref{4_2_comparison}, which illustrates a different phenomenon represented by the pairs symbolized by squares. The BET diagnosis (left panel) and the SCAN-B scatter (middle panel) plots reveal a data threshold issue in the SCAN-B data set. The corresponding TCGA scatter plot (right panel) of this same pair does not have this issue. This threshold effect apparently is caused by missing values in the SCAN-B data set being replaced by the minimum of their values. This same phenomenon occurred for each of the pairs represented by squares in Figure \ref{4_2_comparison}. As discussed in Section \ref{3.1}, we recommend handling such threshold data by jittering.

\begin{figure}[]
\centering
{%
\resizebox*{4cm}{!}{\includegraphics[width=0.5\textwidth,height=0.5\textheight,keepaspectratio]{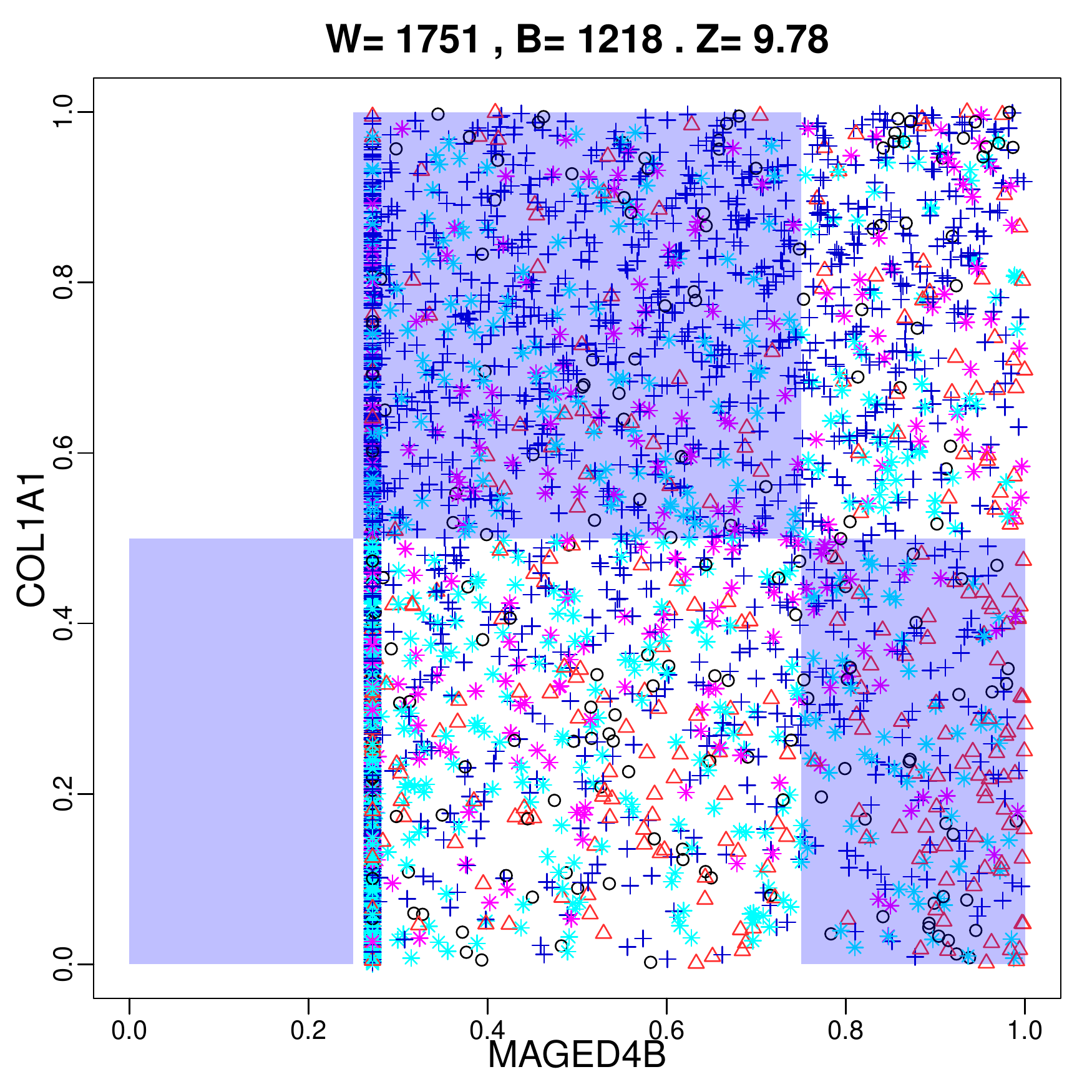}}}\hspace{5pt}
{%
\resizebox*{4cm}{!}{\includegraphics[width=0.5\textwidth,height=0.5\textheight,keepaspectratio]{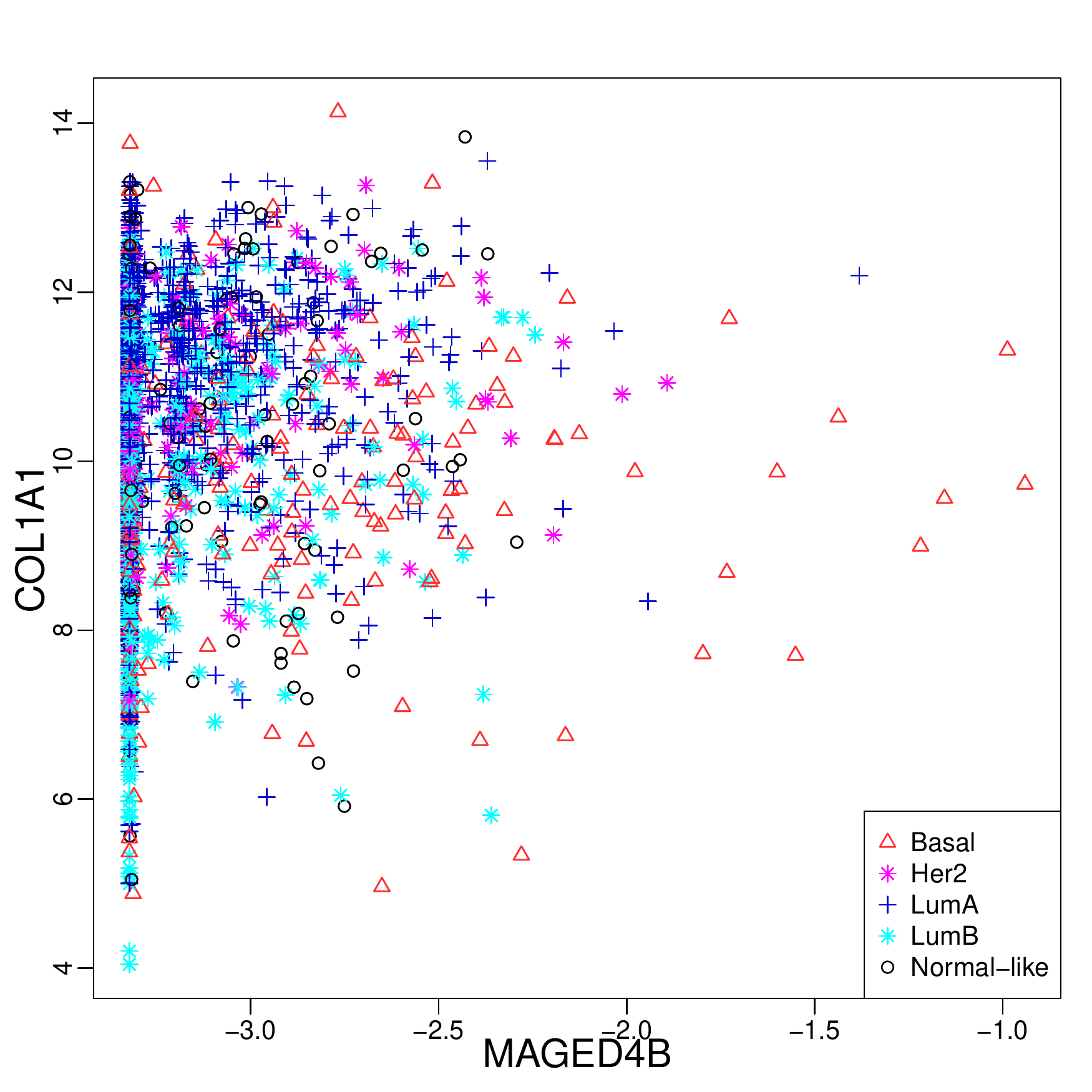}}}\hspace{5pt}
{%
\resizebox*{4cm}{!}{\includegraphics[width=0.5\textwidth,height=0.5\textheight,keepaspectratio]{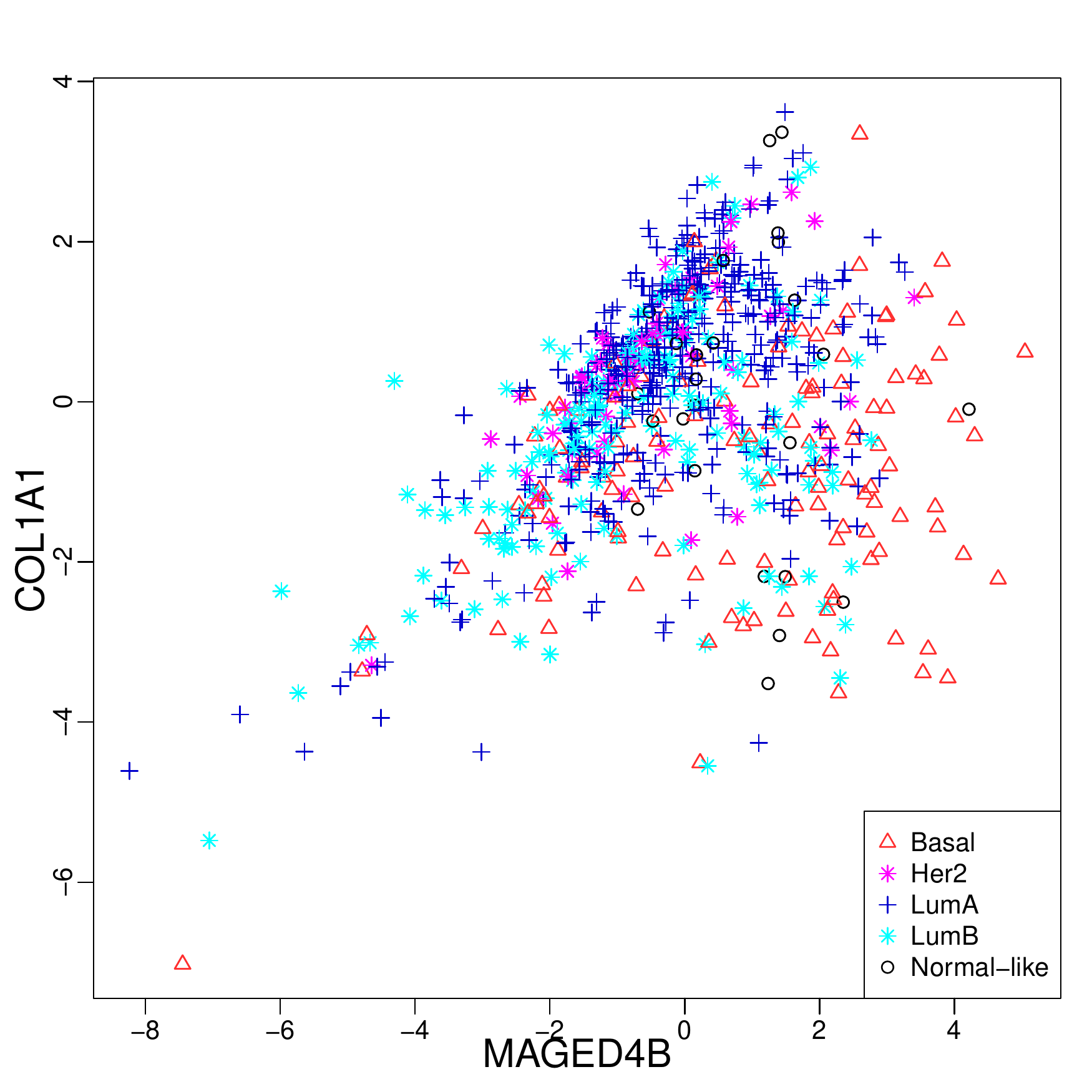}}}
\caption{Left: BET diagnosis plot for a pair of genes in the SCAN-B that is shown as a dark square in Figure \ref{4_2_comparison} revealing a data threshold issue; Middle: The scatter plot of the same two genes in the original SCAN-B scale; Left: The scatter plot of the same two genes in the normalized TCGA count scale. This example represents a type of gene pairs which tend to be less significant in the SCAN-B data set.} \label{4_2_example_1} \label{4_2_example_1}
\end{figure}

\subsection{Lack of Reproducibility of the Bimodal Pattern}\label{lack_rep}

\begin{figure}[]
\centering
{%
\resizebox*{4cm}{!}{\includegraphics[width=0.5\textwidth,height=0.5\textheight,keepaspectratio]{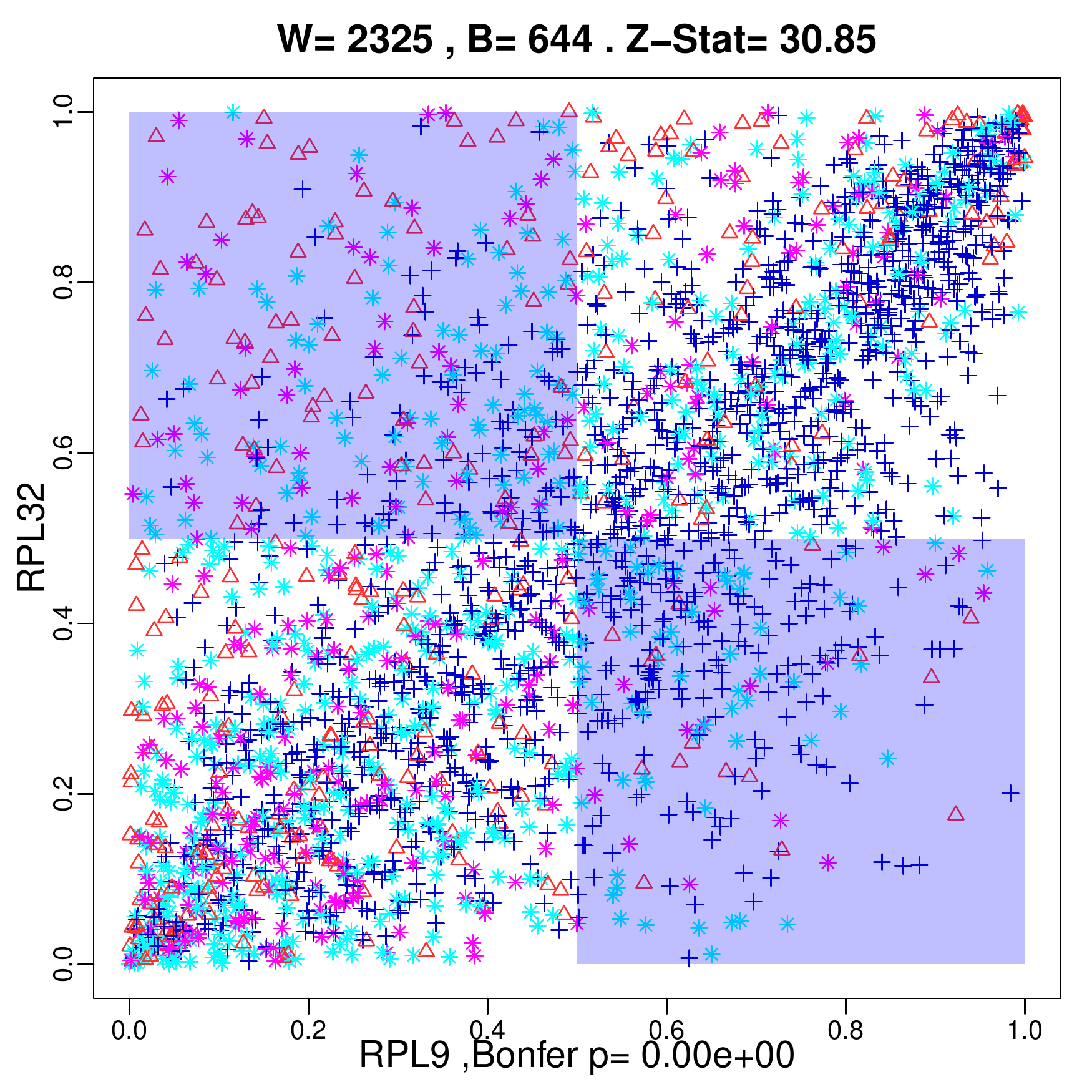}}}\hspace{5pt}
{%
\resizebox*{4cm}{!}{\includegraphics[width=0.5\textwidth,height=0.5\textheight,keepaspectratio]{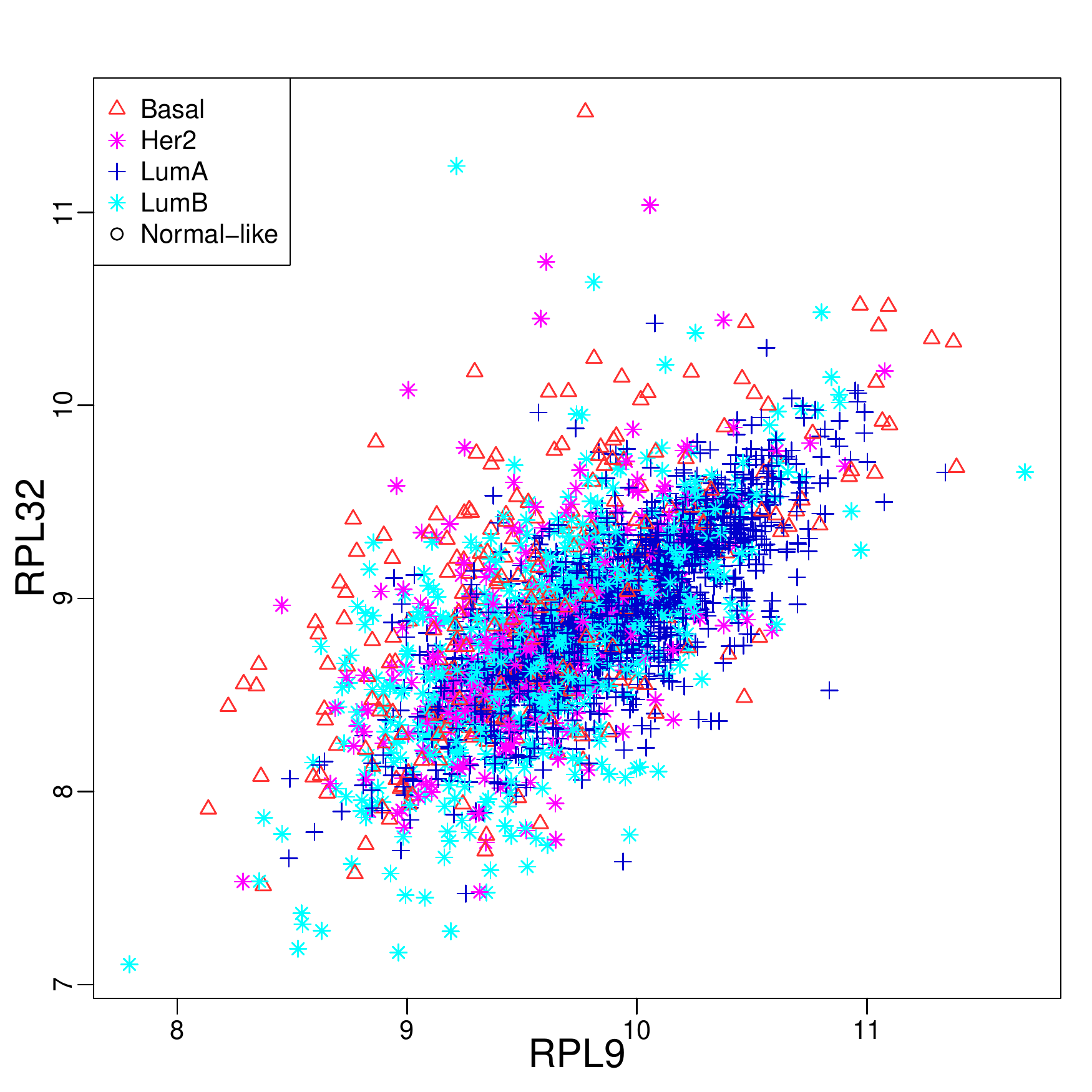}}}\hspace{5pt}
{%
\resizebox*{4cm}{!}{\includegraphics[width=0.5\textwidth,height=0.5\textheight,keepaspectratio]{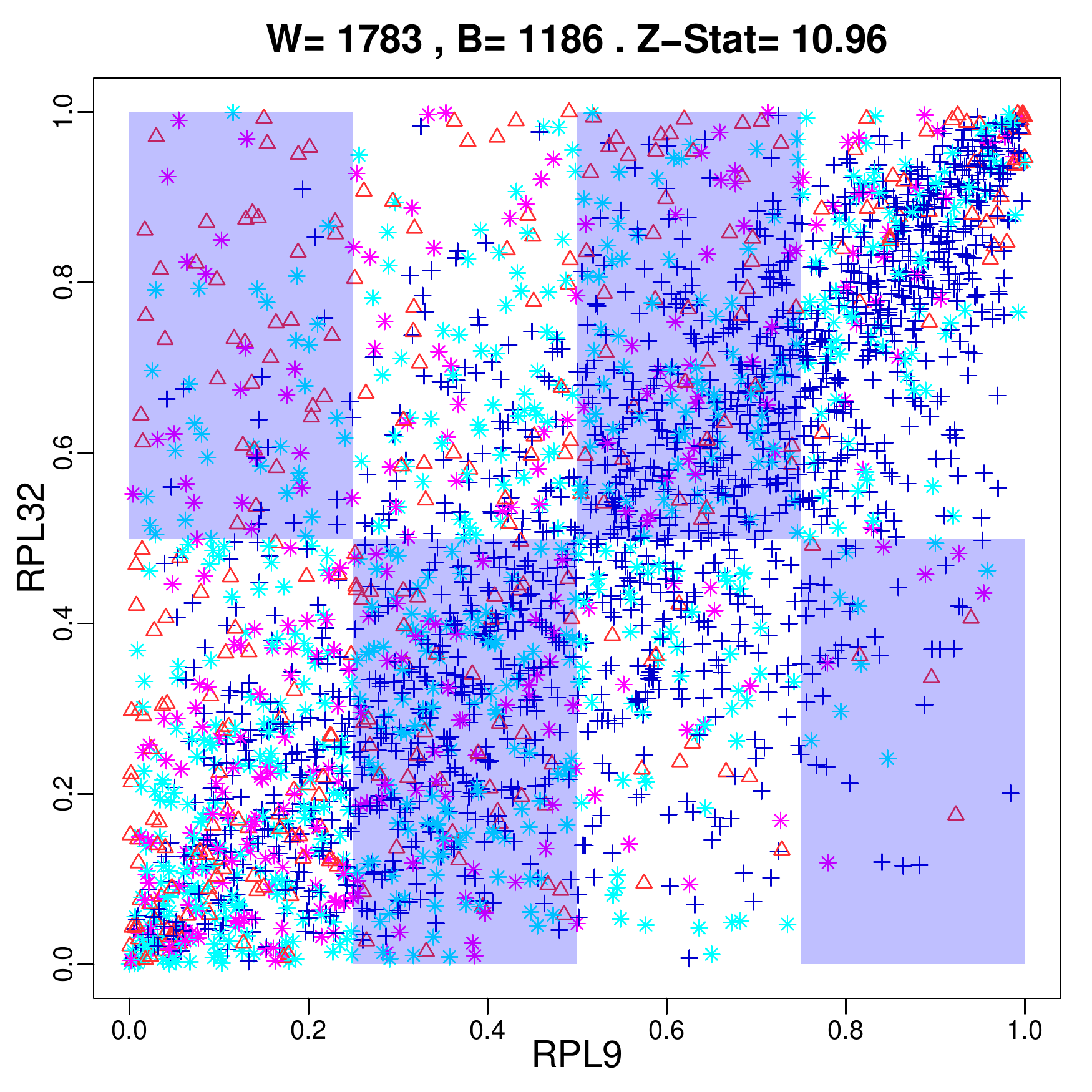}}}
\caption{Left: BET diagnosis plot of the most significant BID ($A_1B_1$) for RPL9 and RPL32 in the SCAN-B data set; Middle: The scatter plot of the same two genes in the original SCAN-B scale; Right: BET diagnosis plot of the W BID for the same two genes. This shows the bimodal pattern observed for RPL9 in TCGA data is not biologically reproducible.} \label{4_1_example_1}
\end{figure}

Here we study the pair of genes that gives the strongest bimodal pattern signal in TCGA data set, which are RPL9 and RPL32 as shown in Figure \ref{3_5_example}. We rerun BET on this pair of genes in the SCAN-B data, and the strongest BID for this pair is linear, as shown in the left panel of Figure \ref{4_1_example_1}. Both the BET diagnosis (left panel) and the scatter (middle panel) plots show a relatively standard positively correlated linear dependence between RPL9 and RPL32. The right panel of Figure \ref{4_1_example_1} gives the BET diagnosis plot for the W BID for this pair, which is much less significant than the linear pattern ($z = 30.85$ vs. $z = 10.96$). This shows that the surprising bimodal dependence observed in Figure \ref{3_5_example} is not biologically reproducible. Instead, it seems to be a processing artifact. As noted above, deeper investigation of the artifact is given in Supplement "Deeper Investigation of the Bimodal Structure of RPL9".

\section{Conclusion}\label{5}

TCGA gene expression data set is an important genomics data resource that shows many dependence patterns among genes, especially some interesting nonlinear dependence patterns. We use the computationally fast and powerful dependence testing method called BET to discover significant nonlinear dependence relationships in various contexts using the breast cancer subtypes information. We find that some interesting nonlinear dependence patterns are explained biologically by the mixture of the given breast cancer subtype distributions, such as the Mixture pattern for the context of Five Subtypes. Some relationships motivate further biological work, such as the Mixture pattern for the LumA only context. We also investigate the reproducibility of these results using an independent genomics data set. This shows that the mixture pattern is reproducible while the bimodal pattern related to the gene RPL9 is not and is apparently caused by some preprocessing steps.


\section{Acknowledgement}
The results published here are in whole or part based upon data from the Cancer Genome Atlas managed by the NCI and NHGRI (dbGaP accession phs000178).

Xiang's research was supported by SAMSI, NIH/NIAMS Grants P30AR072580 and R21AR074685, and DMS-2152289 from NSF.

Perou's research was supported by NCI Breast SPORE program P50-CA58223 and U01CA238475-01.

Zhang's research was partially supported by DMS-1613112, IIS-1633212, DMS-1916237 and DMS-2152289 from NSF.

Marron’s research was supported by NSF Grants IIS-1633074 and DMS-2113404.

\bibliographystyle{plainnat}
\bibliography{reference}

\end{document}


\maketitle

\section{Introduction}

In Section 3.4, we find that ZDHHC2 has the Mixture (the parabolic BID $A_1A_2B_1$) dependence with many genes in the LumA only context in TCGA dataset. As shown in the top pair in Section 3.4, ZDHHC2 plays the bifurcating role in the dependence with each of these other genes. In this supplement, we show two more examples with ZDHHC2 on varying axes. These examples show how ZDHHC2 bifurcates the entire group of observations. They also illustrate the reflection property of BET.

\begin{figure}[H]
\centering
{%
\resizebox*{5.2cm}{!}{\includegraphics[width=1\textwidth,height=1\textheight,keepaspectratio]{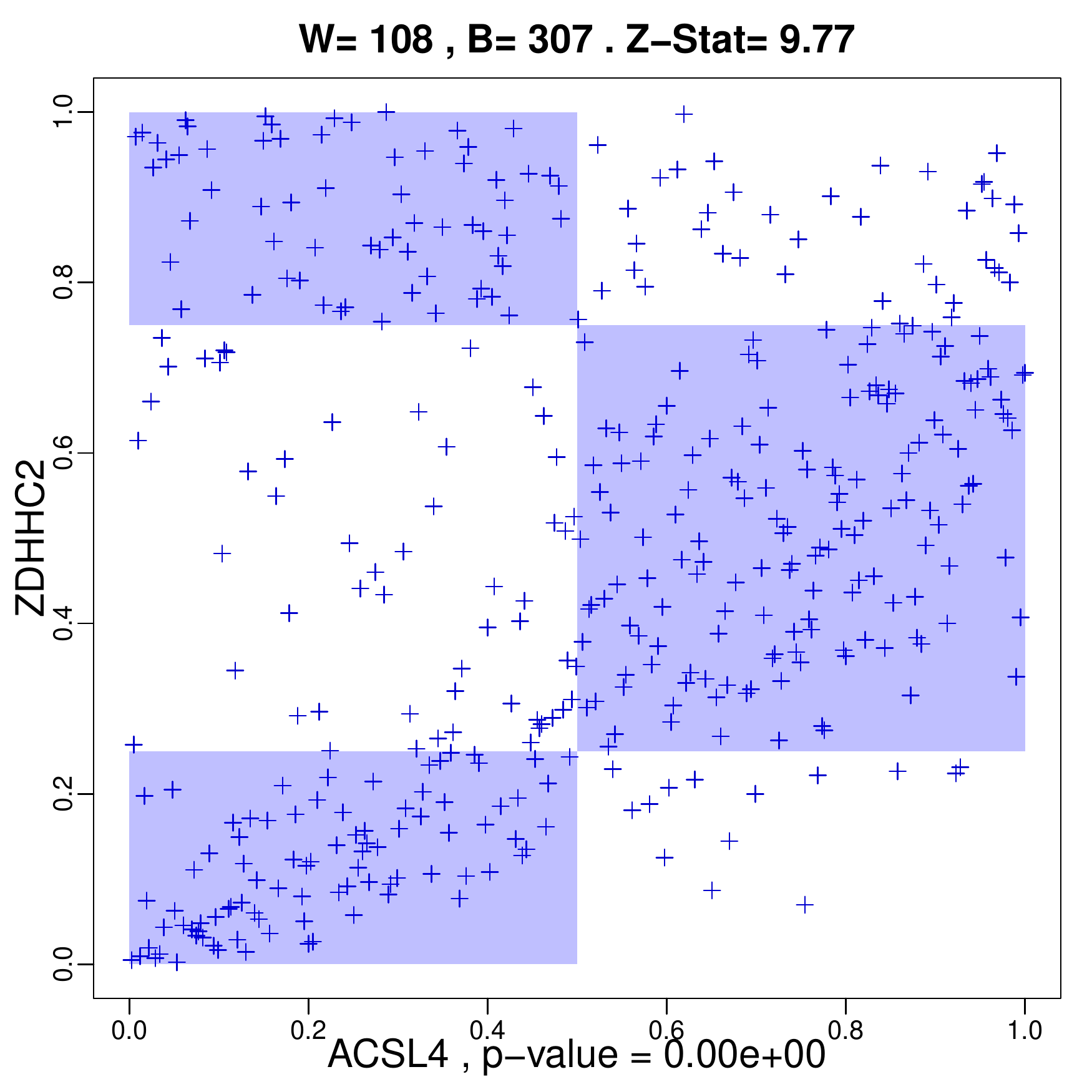}}}\hspace{5pt}
{%
\resizebox*{5.2cm}{!}{\includegraphics[width=1\textwidth,height=1\textheight,keepaspectratio]{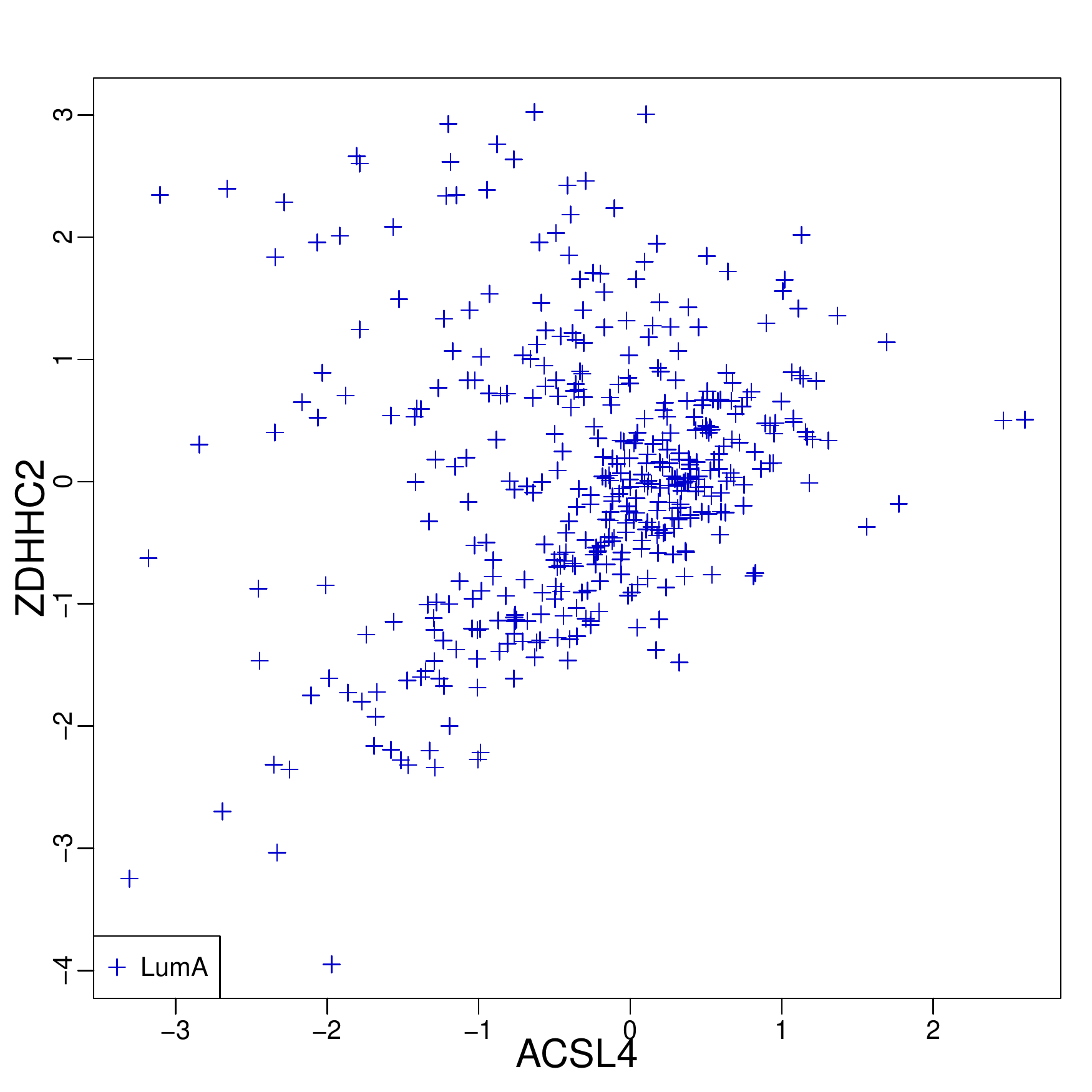}}}
\caption{Left: BET diagnosis plot for a pair of genes with ZDHHC2 in the vertical axis; Right: The scatter plot of the same two genes in the normalized count scale, showing a bifurcation in the same spirit of that in Figure 7.} 
\end{figure}

\begin{figure}[H]
\centering
{%
\resizebox*{5.2cm}{!}{\includegraphics[width=1\textwidth,height=1\textheight,keepaspectratio]{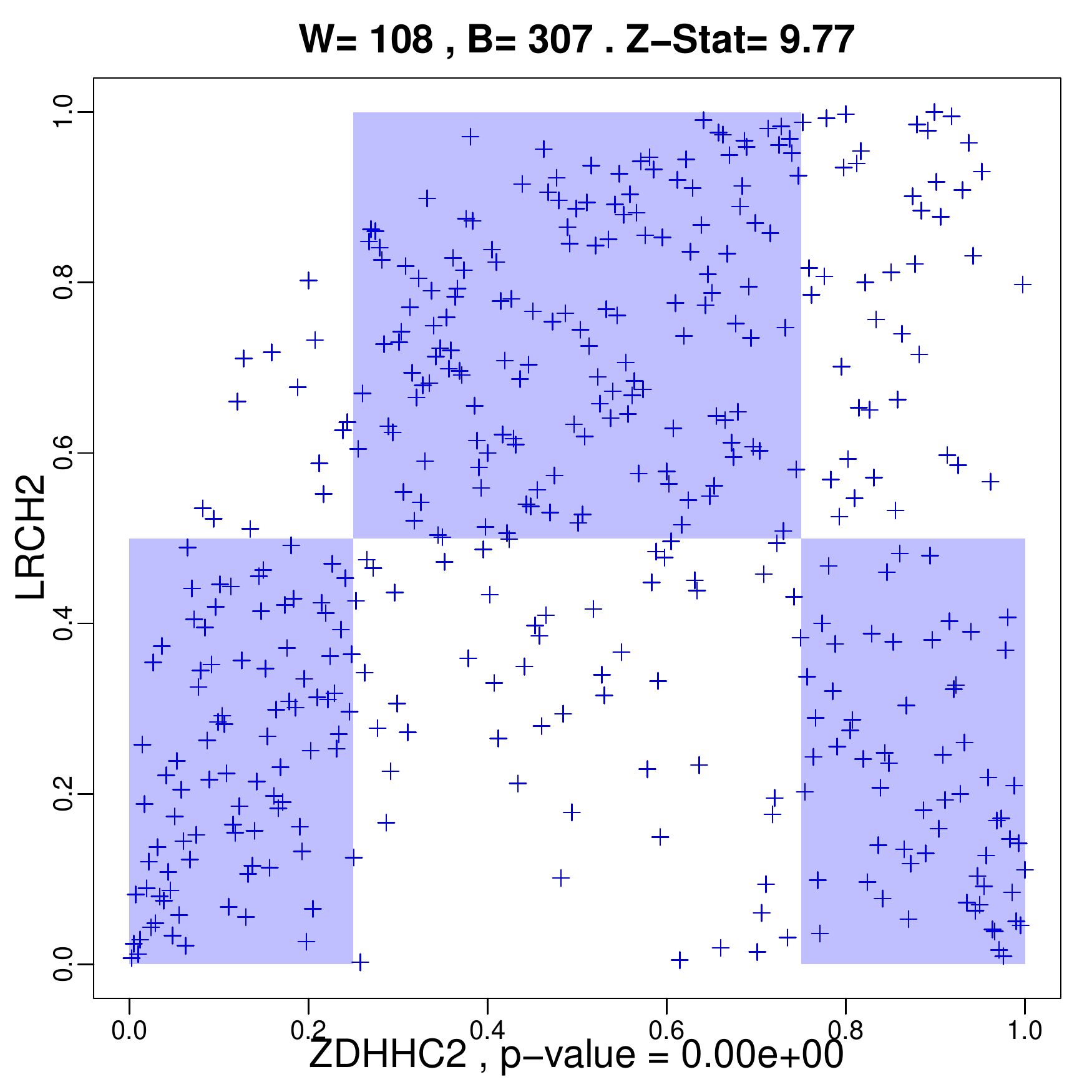}}}\hspace{5pt}
{%
\resizebox*{5.2cm}{!}{\includegraphics[width=1\textwidth,height=1\textheight,keepaspectratio]{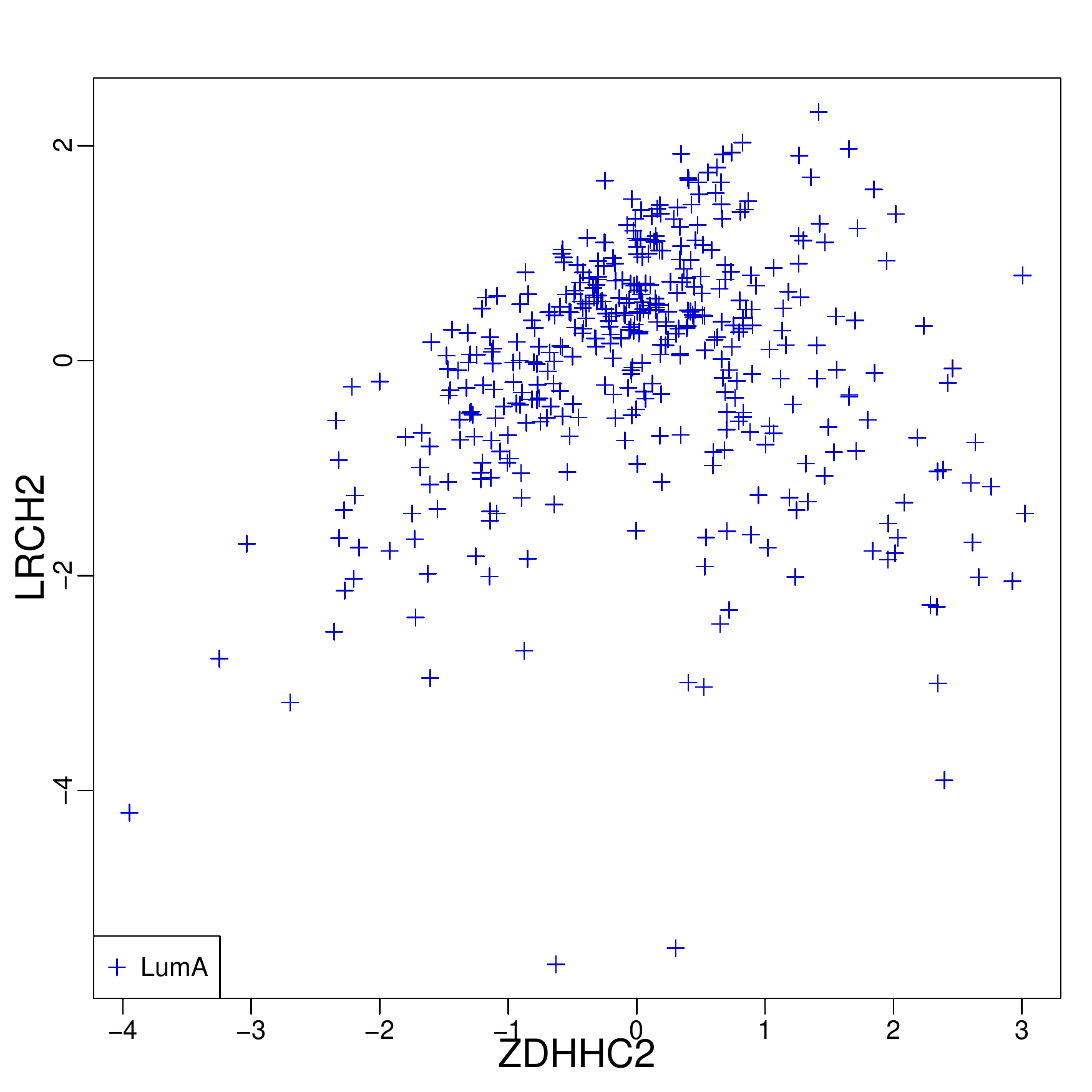}}}
\caption{Left: BET diagnosis plot for a pair of genes with ZDHHC2 in the horizontal axis; Right: The scatter plot of the same two genes in the normalized count scale, conveys a similar main lesson.} 
\end{figure}


\maketitle

\section{Introduction}

\begin{figure}[!htbp]
\centering{
\includegraphics[width=0.85\textwidth,height=0.75\textheight,keepaspectratio]{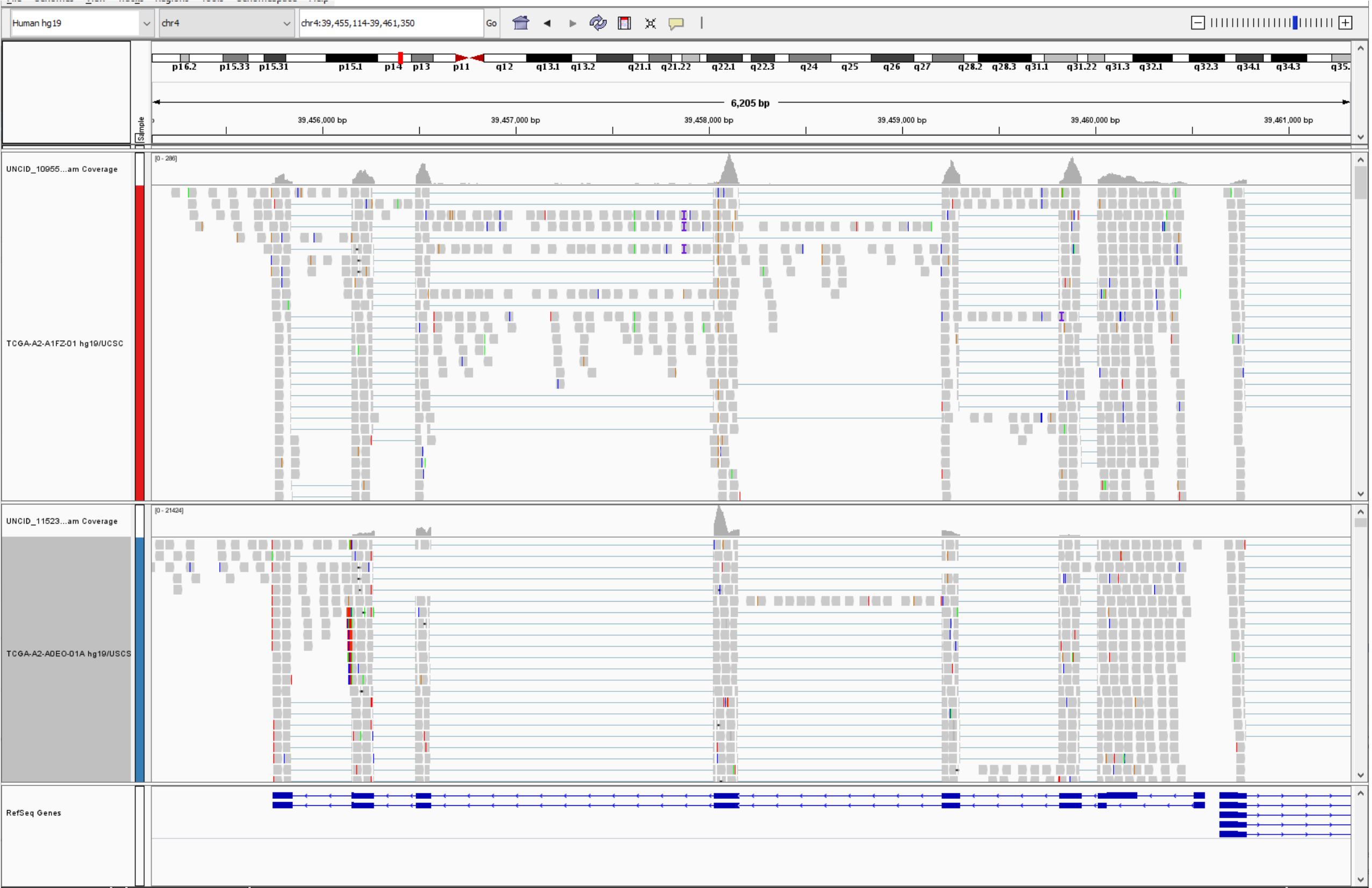}
\caption{The hg19 mapping of the gene RPL9 from two samples: the top sample is from the left hand cluster (low expression) in Figure~10 of the main text and the bottom sample is from the right hand cluster. 
}
\label{hg19}} 
\end{figure}

This section contains a deeper investigation of the surprising bimodal distribution of expression of the gene RPL9 discussed in Sections 3.6 and 4.2.  This is an artifact of the original gene mapping algorithm, based on the relatively early hg19 annotation of the human genome.  Our knowledge of the human genome and gene definitions have improved over time, particularly in regions with high homology and DNA similarity to other parts of the genome. RPL9 is particularly challenging due to many processed pseudogenes that are found on multiple different chromosome regions. This leads to issues with appropriate mapping of reads to the correct location, particularly in short read sequencing where there will be higher overlap across the genome.  In the original TCGA, reads were mapped to hg19 and genes were quantified using RSEM to the set of UCSC Known Genes of around 20,000 RNAs. TCGA used a polyA sequencing approach that selects for mature RNAs that have already spliced out intronic regions. Therefore, when mapping to the genome, reads will only align to exonic regions and can be split across exons in a genomic view.
These effects are illustrated in Figure \ref{hg19}, which includes two examples (i.e.  breast cancer cases) one with low expression (top panel) and one with high expression (bottom panel). This shows the poor alignment to the exon regions in hg19 (indicated using blue boxes at the bottom of the panel). The exon boundaries are not sharp, and there are few split reads from exon to exon (thin blue horizontal lines between reads/boxes). Furthermore, many reads align to intronic regions (between blue boxes).  These poor alignments were observed for cases with both low RPL9 expression (top panel) and high RPL9 expression (bottom panel).

\begin{figure}[H]
\centering{
\includegraphics[width=0.85\textwidth,height=0.75\textheight,keepaspectratio]{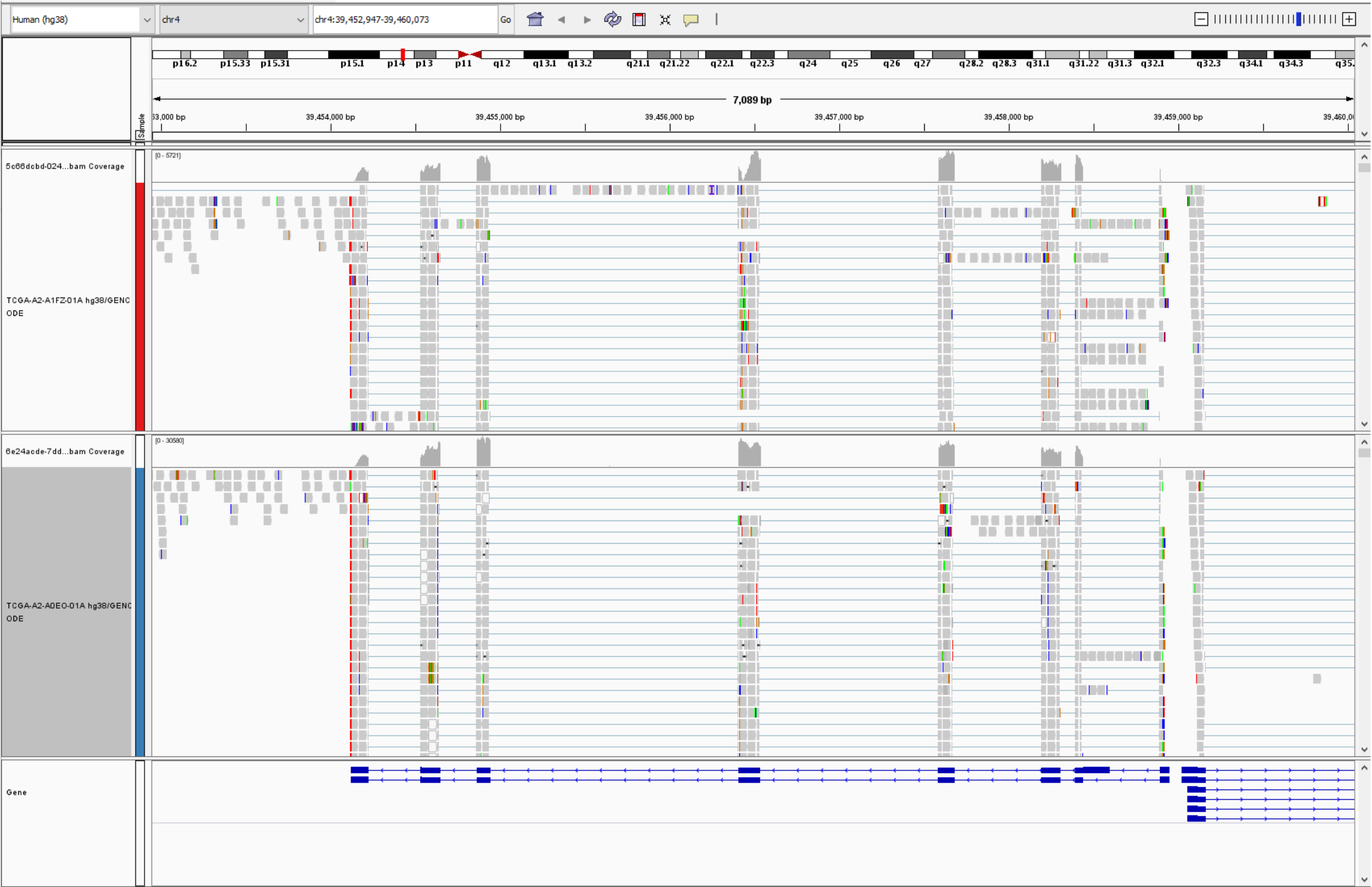}
\caption{
The hg38 mapping of RPL9 for the same two samples (low expression top, high expression bottom). This shows much more consistent mapping, eliminating the spurious bimodality observed in Figure~10.
}
\label{hg38}} 
\end{figure}

These artifacts were not present in the SCAN-B data, because it used the hg38 genome and updated gene annotation and therefore had improved mapping and gene quantification. This improved mapping, for these same samples, is shown in Figure \ref{hg38}. When the same data were aligned to an updated version of the genome hg38, these alignment artifacts were drastically reduced leading to better estimates of RPL9 gene expression.  As seen in Figure \ref{hg38}, there are cleaner exon boundaries, more split reads from exon to exon (thin blue lines), and reduced intronic read mapping (spaces between blue boxes).

\maketitle

In Section 3.3, we use a network connection plot in Figure 6 to analyze
the relationships between pairs of 200 most significant genes with respect to only the same BID $A_1B_1B_2$. Here we use a network connection plot to show the relationships among the 200 most significant genes over all five nonlinear BIDs in the Five Subtype context of the top row in Figure 4. Nodes represent genes. Each edge highlights the most significant BET BID for each pair of genes: grey edges represent the Parabolic BID ($A_1A_2B_1$ and $A_1B_1B_2$) as discussed in Section 3.3 and 3.4; blue edges represent the W BID ($A_2B_1$ and $A_1B_2$)  as discussed in Section 3.6; green edges represent BID $A_1A_2B_1B_2$ as discussed in Section 3.5. There is no overlap of edges, because only the most significant BID is recorded for each pair. Genes at the center of some visually important communities are labeled. Notice that most of the top significant pairs of genes are significant with respect to the Parabolic BID (grey), so this figure is quite similar to Figure 6. All pairs in the RPL9 Gene community (blue) are with respect to the W BID, which are not reproducible as shown in Section 4.2 and Supplement D. A few pairs are with respect to BID $A_1A_2B_1B_2$.

\begin{figure}[]
\centering
\includegraphics[width=1\textwidth,height=1\textheight,keepaspectratio]{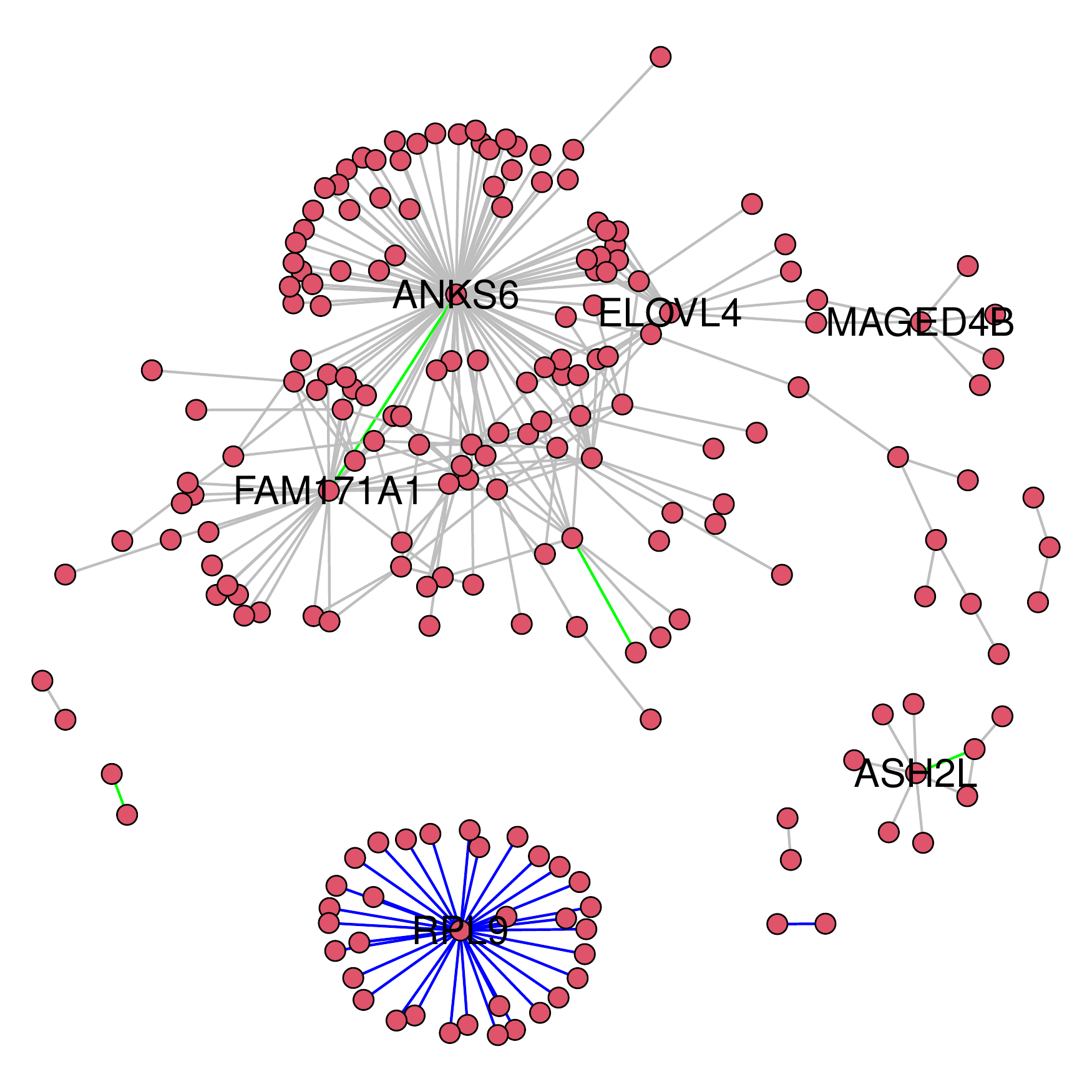}
\caption{Network connection plot of 200 most significant genes (nodes) with 270 edges for five nonlinear BIDs. Each node represents one gene and each edge represents a significant dependence between those genes. Color represents the type of BID. The large community illustrated there are many genes having significant dependence pattern with the center gene, such as ANKS6. Discovered communities are very similar to those in Figure 6.}
\label{allNonlinearBIDs_network}
\end{figure}